\documentstyle[12pt]{article}
\newlength{\pubnumber} 

\catcode`\@=11
\@addtoreset{equation}{section}

\def\section{\@startsection{section}{1}{\z@}{3.5ex plus 1ex minus .2ex}
 {2.3ex plus .2ex}{\large\bf}}
\def\subsection{\@startsection{subsection}{2}{\z@}{2.3ex plus .2ex}
 {2.3ex plus .2ex}{\bf}}

\begin{document}

\begin{titlepage}
\samepage{
\setcounter{page}{1}
\rightline{McGill/93-01}
\rightline{\tt hep-th/9305094}
\rightline{May 1993}
\vfill
\begin{center}
 {\Large \bf The Worldsheet Conformal Field Theory\\
  of the Fractional Superstring\\}
\vfill
 {\large Keith R. Dienes\footnote{E-mail address:
  dien@hep.physics.mcgill.ca.}\\}
\vspace{.10in}
 {\it  Department of Physics, McGill University\\
  3600 University St., Montr\'eal, Qu\'ebec~H3A-2T8~~Canada\\}
\end{center}
\vfill
\begin{abstract}
 {\rm Two of the important unresolved issues concerning fractional superstrings
 have been the appearance of new massive sectors whose spacetime statistics
 properties are unclear, and the appearance of new types of ``internal
 projections'' which alter or deform the worldsheet conformal field theory
 in a highly non-trivial manner.  In this paper we provide a systematic
 analysis of these two connected issues, and explicitly map out the
 effective post-projection worldsheet theories for each of the
 fractional-superstring sectors.  In this way we determine their central
 charges, highest weights, fusion rules, and characters, and find that
 these theories are isomorphic to those of free worldsheet bosons and fermions.
 We also analyze the recently-discovered parafermionic ``twist current''
 which has been shown to play an important role in reorganizing the
 fractional-superstring Fock space, and find that this current can be
 expressed directly in terms of the primary fields of the post-projection
 theory.  This thereby enables us to deduce some of the spacetime statistics
 properties of the surviving states.}
\end{abstract}
\vfill}
\end{titlepage}

\setcounter{footnote}{0}

\def\beq{\begin{equation}}
\def\eeq{\end{equation}}
\def\beqn{\begin{eqnarray}}
\def\eeqn{\end{eqnarray}}
\def\calZ{{\cal Z}}
\def\bZ{{\bf Z}}
\def\half{{\textstyle {1\over2}}}
\def\quarter{{\textstyle {1\over4}}}
\def\quart{{\textstyle {1\over4}}}
\def\eig{{\textstyle {1\over8}}}
\def\smallfrac#1#2{{ \textstyle{ {#1}\over{#2} }}}
\def\bone{{\bf 1}}
\def\ie{{\it i.e.}}
\def\eg{{\it e.g.}}
\def\tautwo{{ \tau_2}}
\def\qbar{{  \overline{q} }}
\def\thetafour{{ \vartheta_4 }}
\def\thetatwo{{ \vartheta_2 }}
\def\thetathree{{ \vartheta_3 }}
\def\tw { \tilde }
\def\current{{\Psi}}
\def\currentfour{{\Psi_4}}
\def\currenteight{{\Psi_8}}
\def\IC{\relax\hbox{$\inbar\kern-.3em{\rm C}$}}
\def\IQ{\relax\hbox{$\inbar\kern-.3em{\rm Q}$}}
\def\IR{\relax{\rm I\kern-.18em R}}
 \font\cmss=cmss10 \font\cmsss=cmss10 at 7pt
\def\IZ{\relax\ifmmode\mathchoice
 {\hbox{\cmss Z\kern-.4em Z}}{\hbox{\cmss Z\kern-.4em Z}}
 {\lower.9pt\hbox{\cmsss Z\kern-.4em Z}}
 {\lower1.2pt\hbox{\cmsss Z\kern-.4em Z}}\else{\cmss Z\kern-.4em Z}\fi}

\hyphenation{pa-ra-fer-mion pa-ra-fer-mion-ic pa-ra-fer-mions }
\hyphenation{su-per-string frac-tion-ally su-per-re-pa-ra-met-ri-za-tion}
\hyphenation{su-per-sym-met-ric frac-tion-ally-su-per-sym-met-ric}
\hyphenation{space-time-super-sym-met-ric fer-mi-on}
\hyphenation{Ne-veu-Schwarz-like Ramond-like}


\setcounter{footnote}{0}
\section{Introduction}

Over the past two years there has been considerable
activity in a possible new class of string theories
known as {\it fractional superstrings} [1--7]:
these are non-trivial generalizations of superstrings and heterotic strings,
and have the important property that
their critical spacetime dimensions are less than ten.
This reduction in the critical dimension
is accomplished by replacing the worldsheet supersymmetry
of the traditional superstring or heterotic string
by a $K$-fractional supersymmetry:
such symmetries relate worldsheet bosons not to worldsheet fermions,
but rather to worldsheet $\IZ_K$ parafermions
of fractional spin $2/(K+2)$.  One then finds that the corresponding
critical spacetime dimension of the theory is given by
\beq
   D_c ~=~ 2 ~+~ {{16}\over K} ~,~~~~K\geq 2~.
\label{kcritdimensions}
\eeq
Thus while the choice $K=2$ reproduces the ordinary $D_c=10$ superstring
(with $\IZ_2$ ``parafermions'' reducing to ordinary Majorana fermions),
the choices $K=4,8,$ and $16$ yield new theories with $D_c=6,4,$ and $3$
respectively.

For $K>2$, however,
these new worldsheet theories are substantially more difficult
to study than those of ordinary superstrings,
since the fundamental worldsheet {\it fractional}\/ superconformal algebra is
necessarily non-local for $K>2$,
with branch {\it cuts}\/ (rather than poles) appearing in the various
parafermionic operator product expansions.  This implies, for example,
that the corresponding mode algebras involve neither commutation nor
anti-commutation relations, but rather the more difficult {\it generalized
commutation relations}\/;  moreover, in many cases this also gives rise to
non-trivial braiding relations for the underlying conformal fields.

It is primarily due to such non-linear complications on the worldsheet that
fractional superstrings appear to exhibit qualitatively new features
in spacetime, as compared to the usual superstrings and heterotic strings.
Understanding these new features
is thus of paramount importance, not only for demonstrating
the internal consistency of the fractional superstring,
but also as a means of shedding further light on
the general but as yet poorly understood relationship between worldsheet
string symmetries and spacetime physics.
Therefore, in order to gain some insight into their spacetime physics,
fractional superstrings have
been studied from a variety of different perspectives;
a non-technical review is given in \cite{Review}.
One straightforward approach in analyzing these theories
is essentially ``bottom-up'', and attempts to derive the resulting
spacetime physics of the fractional superstring (including the
spacetime partition functions) by starting from the original worldsheet
parafermionic theory described above
and constructing the complete resulting Fock space of states.
While this approach has proven successful for
understanding the properties of certain low-lying states \cite{ALT},
difficulties arise at the higher mass levels of the fractional-superstring
spectra.  Another approach is essentially ``top-down'', and starts by
constructing the unique modular-invariant partition functions of the forms
that these theories must have;  by analyzing these partition functions
in great detail one can then obtain information about the underlying
worldsheet physics from which they might ultimately be derived.
This approach has also been successful to a large extent
\cite{DT,ADT,FL,CR,AD}, and has shed light on many of the intrinsically new
features of the fractional superstrings.  Both complimentary approaches
have uncovered but left unresolved, however, two crucial aspects of these
new string theories which have no analogues in the traditional
super- or heterotic string theories:  (1) the appearance of
extra unusual massive sectors (the so-called ``$B$-sectors'') which contain
spacetime particles whose physical roles are unclear;  and (2) the appearance
of new so-called ``internal projections'' which, unlike the traditional GSO
projection, remove a sufficient number of states from the fractional
superstring spectra to actually diminish the effective central charges of
these theories.  In fact, as we will see, these internal projections
have the net effect of changing or deforming the underlying
worldsheet parafermionic conformal field theories (CFT's)
upon which the fractional superstring is originally built,
with the surviving states filling out the Fock spaces corresponding
to new worldsheet CFT's whose properties and worldsheet
representations are as yet unknown.
It is towards developing a better understanding of these
connected issues that this paper is addressed, for these are the features
which ultimately reflect the non-linearities of the worldsheet theory
and which contain much of the new physics of the fractional
superstring.  While we certainly do not have complete
resolutions to these puzzles, our results provide
the first clues concerning both the effective worldsheet
conformal field theory which survives the internal projections
in each of the fractional superstring sectors,
and the spacetime statistics properties of the surviving states.

In particular, our main results may be briefly summarized as follows.
Whereas the original worldsheet conformal field theory of the $K$-fractional
superstring in light-cone gauge has central charge $c=48/(K+2)$ and consists
of a tensor product of $D_c-2=16/K$ coordinate bosons tensored together
with $16/K$ copies of the $\IZ_K$ parafermion theory,
\beq
   {\rm CFT}~=~ \left(
  {\mathop \otimes _{\mu=1}^{D_c-2=16/K}}   X^\mu \right) ~\otimes~
 \left\lbrace {\mathop \otimes _{\mu=1}^{D_c-2=16/K}}
     (\IZ_K ~{\rm PF})^\mu \right\rbrace~,
\label{originalCFT}
\eeq
we find that the internal projections reduce this theory down to
the smaller $c=24/K$ conformal field theory
\beq
   {\rm new~CFT}~=~ \left(
  {\mathop \otimes _{\mu=1}^{D_c-2=16/K}}   X^\mu \right) ~\otimes~
   \left( c={8\over K} ~{\rm theory}\right)
\label{finalCFT}
\eeq
where this $c=8/K$ component is a certain {\it non}\/-tensor-product
theory whose central charge, highest weights, fusion rules, and characters
render it isomorphic to
a tensor product of $8/K$ bosons compactified on circles of certain radii
(or to a single $c=1/2$ Ising model in the $K=16$ case).
Specifically, this means for each relevant value of $K$,
these $c=8/K$ post-projection theories
have the same central charges, highest weights, fusion
rules, and characters as those of $8/K$ free compactified bosons ---
even though (as we shall demonstrate) the post-projection CFT's
cannot be represented in this simple manner.
Moreover, we find that this ``isomorphism''
between our post-projection theories and these compactified-boson theories
holds for {\it all}\/ sectors of the fractional superstring,
including the extra massive $B$-sectors,
and that the only difference between these new massive sectors and
the more traditional Neveu-Schwarz-like and Ramond-like sectors
is an apparent change in the compactification radius of the
bosons in the isomorphic theory.
We also analyze the so-called parafermionic ``twist current'' which plays
a crucial role in reorganizing the Fock space during the internal
projections, and surprisingly find that this current can be represented
as a certain primary field in the resulting {\it post}\/-projection theory.
This then enables us to identify some of the spacetime statistics
properties of the states surviving the internal projections,
and to express all of our results
in terms of an effective compactification lattice.

This paper is organized as follows.  In Section 2 we first provide a
non-technical overview of these two fundamental issues which confront the
fractional superstring, followed by a more technical introduction
in which we discuss the appearance of these new features
and summarize some recent results upon which our work is based.
We then proceed in Section 3 to examine the effects of these new internal
projections on the better-understood Ramond- and Neveu-Schwarz-like sectors
of the fractional superstring (the so-called ``$A$-sectors''), and deduce
many of the properties of the new $A$-sector conformal field theories which
emerge after these internal projections have acted.  In particular, we find
that we are able to explicitly construct a mapping between the sectors of
the pre-projection and post-projection conformal field theories in the
$A$-sectors, and this allows us to obtain a set of minimal constraints
({\it i.e.}, the central charges, highest weights, fusion rules, and
characters) that these post-projection conformal field theories must satisfy.
In Section 4 we repeat our analysis for the $B$-sectors, and in Section 5 we
demonstrate that the post-projection CFT's for both the $A$- and $B$-sectors
closely resemble those of free worldsheet compactified bosons.  We then
rewrite our results in such a way that this isomorphism is manifest, and
in Section 6 we use this reformulation to analyze the parafermionic ``twist
current''.  In particular, we will find that we can express this twist
current directly in language of the isomorphic compactified-boson theory,
and this in turn will enable us to understand some of the spacetime
statistics properties of the fractional-superstring sectors.
We then close in Section 7 with a summary of our results, and with
comments regarding the possibility of constructing a unified worldsheet
conformal field theory capable of simultaneously describing {\it all}\/ of
the post-projection fractional-superstring sectors.

Our primary motivation is to discover various properties of the
post-projection CFT's in both the $A$- and $B$-sectors, and as such our goals
are two-fold.  First, we seek to demonstrate that the new $B$-sectors are
consistent with the internal projections, and from the vantage
point of the {\it post}\/-projection theory,
we will see that these new sectors closely resemble
(rather than differ from) the $A$-sectors
whose properties are better understood.  This in itself should be
of great importance in ultimately demonstrating the consistency of
these new string theories.  We will also find, however, that we cannot yet
construct suitable representations for these post-projection CFT's
in either the $A$- or $B$-sectors, for some technical issues having to do with
spacetime statistics remain as yet unresolved.
Our second goal, therefore, is to somewhat broadly set forth a set of
minimal conditions that these CFT's and their appropriate representations
must ultimately satisfy.  In doing so, we will be following almost
exclusively the ``top-down'' approach discussed earlier, exploiting the
partition-function evidence as much as possible in order to provide insight
into these post-projection CFT's.   Our results can then hopefully
serve as a guide in any future ``bottom-up'' construction.

\section{Massive Sectors and Internal Projections}
\setcounter{footnote}{0}

In Sects.~2.1 and 2.2 we first provide a non-technical overview
of the two fundamental issues which currently confront the fractional
superstring.  Sect.~2.3 then contains a more technical
review of fractional superstrings and their constituent
parafermion theories.

\subsection{Massive Sectors}

Fractional superstrings (like ordinary bosonic strings,
superstrings, and heterotic strings) have spacetime particle
spectra consisting of various infinite towers of states:  each tower
represents the Fock space of states built upon
a unique vacuum state, and each of these various vacuum states corresponds
to a certain primary field
in the underlying worldsheet conformal field theory.
The spacetime (mass)$^2$ of each vacuum state
is of course related to the highest weight of the corresponding primary field
via $m^2=h-c/24$ where $h$ is this highest weight and where $c$
is the central charge of the underlying conformal field theory;
similarly, the states in each tower have values of $m^2$
differing from those of their vacuum state only by integers.
We will refer to each of these towers as a (conformal
field theory) ``sector'' of the theory,
avoiding the more traditional definition of a string-theory ``sector''
in terms of the toroidal boundary conditions of worldsheet fields.
What will interest us here is the appearance of new massive
fractional-superstring sectors which have no analogues in ordinary
superstrings.

In order to specify the sense in which these sectors are new, it proves
instructive to recall the case of the ordinary superstring in $D=10$.
The underlying light-cone worldsheet conformal field theory
of the usual superstring has central charge $c=12$, and consists of a tensor
product of eight free bosons and eight free Majorana
fermions (each of the latter being equivalent to a $c=1/2$ Ising model).
While each of the eight bosonic CFT's has only one sector (the identity
sector with $h=0$), each Ising-model factor has three sectors
(the identity sector $[\bone]$ with $h=0$,
the fermion sector $[\psi]$ with $h=1/2$, and
the spin-field sector $[\sigma]$ with $h=1/16$).
There are thus, {\it a priori}, a total of ${{10}\choose 2}=45$ possible
sectors in the superstring, where we are not distinguishing
the {\it order}\/ of the eight Ising-model factors.\footnote{
  Note that since this worldsheet theory is a tensor product,
  spacetime Lorentz invariance implies invariance under permutations
  of the transverse dimensions.}
Not all of these potential sectors contribute
spacetime particles to the physical spectrum, however.
As is well-known, the four vacuum states $\bone^7\psi,
\bone^5\psi^3, \bone^3\psi^5$, and $\bone\psi^7$
are the so-called Neveu-Schwarz (NS) vacuum states which contribute
to the superstring spectrum, and the four towers
of states respectively built upon these vacua together
comprise the so-called NS sector of the theory.
All particles in these sectors are spacetime bosons, and have $m^2\geq 0$
with $m^2\in \IZ$.
(By contrast, the five complementary vacuum states
$\bone^8, \bone^6\psi^2,\ldots,\psi^8$, along with their infinite towers
of states, do not appear in the superstring spectrum and are
said to have been removed by the GSO projection.)
Likewise, there is a fifth vacuum state which contributes to the
spectrum of the ordinary superstring:  this is the massless Ramond
vacuum state $\sigma^8$, with all states in the corresponding tower
comprising the so-called Ramond sector of the theory.  All excitations
in this sector are spacetime fermions with $m^2\in \IZ$, and these are
in fact the spacetime superpartners of the particles arising
in the four NS sectors (thus rendering the ordinary superstring
spacetime-supersymmetric).
The crucial observation, however, is that none of the remaining 35 potential
``mixed'' $\bone/\sigma$, $\psi/\sigma$, or $\bone/\psi/\sigma$
vacuum-state combinations contributes to the physical spectrum of
states of the ordinary $D=10$ superstring.

For the more general fractional superstrings with $K>2$, this is no longer
the case:  there are a variety of fundamentally new sectors which contribute
states to the spacetime spectrum and which must therefore be considered.
For each value of $K\geq 2$, the light-cone worldsheet conformal field
theory of the fractional superstring consists of tensor products of
$D_c-2=16/K$
pairs of free bosons and $\IZ_K$ parafermions, and for $K>2$ these $\IZ_K$
parafermion theories contain in principle {\it many}\/ different independent
sectors.   These sectors still fall, however, into certain groups:  the first
group (to be collectively denoted $\lbrace\bone\rbrace$)
contains the higher-$K$ analogues
of the Ising-model sectors $[\bone]$ and $[\psi]$, the second group (to be
collectively denoted $\lbrace \sigma\rbrace$)
contains the analogues of the Ising-model
sector $[\sigma]$, while a third group (to be denoted $\lbrace \phi\rbrace$)
contains those parafermionic
sectors having no analogues at all in the $K=2$ special case.  It is thus
possible to classify the resulting light-cone tensor-product sector
combinations which actually contribute to the spacetime spectrum of the
fractional superstring, distinguishing those which either are or are not
analogues of those encountered in the ordinary superstring.
First, for example, there are various the NS-like combinations
$\lbrace\bone\rbrace^{D_c-2}$:  those which have integer values of $m^2$ again
contribute to the resulting physical spectrum, while the others experience
a GSO projection and do not appear.  Second, there are the
various Ramond-like combinations $\lbrace\sigma\rbrace^{D_c-2}$:  those having
integer values of $m^2$ again contribute to the spacetime spectrum,
while the others are projected away.  These NS and Ramond
sectors, for example, together yield the massless supergravity multiplet
in the fractional superstring,
and are in fact complete superpartners of each other at all mass levels.
These are the so-called ``$A$-sectors'' of the fractional superstring.

There are, however, two other kinds of sectors which contribute
to the fractional-superstring spectrum.  The first (the
so-called ``$B$-sectors'') are
the higher-$K$ analogues of the 35 ``mixed'' superstring sectors:
these all turn out to have the equally-mixed
form $(\lbrace\bone\rbrace\lbrace\sigma\rbrace)^{(D_c-2)/2}$,
and contain only states with masses
$m^2\in \IZ+\half$ with $m^2>0$ (\ie, states at the Planck scale).
The second class consists of sectors (the so-called ``$C$-sectors'')
built exclusively from
the individual parafermionic $[\phi]$ sectors:  these new sectors all take
the form $\lbrace\phi\rbrace^{D_c-2}$, and thus
have no analogues in the ordinary
superstring.  Like the above ``mixed'' sectors, however, they
also {\it a priori}\/ contribute states to the fractional-superstring
spectrum, with masses $m^2\in \IZ+{\textstyle {3\over 4}}$ with $m^2>0$.
Determining the spacetime statistics of the particles in these
new sectors is highly non-trivial,
for the vacuum states upon which these sectors are built are not of the
standard NS or Ramond variety.
These $B$- and $C$-sectors are the ``massive sectors''
whose physical properties we seek to understand.

\subsection{Internal Projections}
\setcounter{footnote}{0}

The second fundamental issue which has remained unresolved concerns the
appearance of new types of ``internal projections'' which,
like the GSO projection discussed above, prevent certain states
in the Fock space of the worldsheet CFT
from appearing in the actual physical spacetime spectrum.
These new internal projections are, however, quite different from
the GSO projections.  As we have seen, the GSO projections
remove states from the physical spectrum only by eliminating the
contributions from entire {\it towers}\/ of states:  any given
tower, including the vacuum state as well the infinite Fock space of
states it generates,
will either fully contribute to the physical spectrum or suffer a complete
GSO projection and not appear at all.
Indeed, such an all-or-nothing, tower-by-tower projection is the
only way in which the resulting spacetime spectrum can still be consistent with
the underlying conformal field theory that gave rise to it, for any
sectors which survive the GSO projection are guaranteed to be
the intact highest-weight sectors of that underlying conformal field theory.

In the fractional superstring theories, however, a new type
of ``internal projection'' appears whose action is far more subtle.
Rather than project out entire towers of states, these new projections
project away only {\it some} of the states in each individual tower,
leaving behind a set of states which clearly {\it cannot}\/
be interpreted as the complete Fock space of the original underlying
worldsheet conformal
field theory.  On the face of it, this would seem to render the
spacetime spectra of the fractional superstrings hopelessly inconsistent
with any underlying worldsheet-theory interpretation.

Remarkably, however, evidence suggests that the residual states which
survive the internal projections in each tower precisely recombine to
fill out the complete Fock space of a {\it different}\/ underlying conformal
field theory.  Thus, whereas the GSO projection merely removed certain
highest-weight sectors of the worldsheet conformal field theory,
these new internal projections appear to actually {\it change the underlying
conformal field theory itself.}~
This turns out to be a profound alteration.
Since the central charges of the new (post-projection)
conformal field theories are smaller than those of the
original tensor-product parafermion theories, these internal projections
must clearly remove exponentially large numbers of states from each
of the mass levels of the original Fock space;
indeed, it is only in this drastic yet highly non-trivial manner
that the fractional-superstring spacetime
spectrum can remain consistent with an underlying
two-dimensional worldsheet theory interpretation.
Such a delicate projection clearly has no analogue in the
ordinary superstring, and perhaps more closely resembles the BRST projection
which enables unitary minimal models to be constructed
from free bosons in the Feigin-Fuchs construction.

Verifying that the internal projection indeed leaves behind a
self-consistent Fock space is of course a difficult task, and
to date the evidence for this has only been obtained via
the ``top-down'' approach --- \ie, through the partition functions and the
implied degeneracies of states.
In fact, it is only in this manner that the internal projections
are evident;  there does not presently exist any formulation
of these internal projection in terms of, for example, a projection
operator constructed out of worldsheet fields.
Therefore, we too shall be forced to follow this ``top-down'' approach,
and indeed we will see that this is sufficient to
determine the central charges, highest weights, fusion rules, and characters
of the light-cone worldsheet conformal field theory
that remains after the internal
projections.  Actually constructing a suitable representation for this
conformal field theory in terms of worldsheet fields remains an open question,
however, and we shall discuss some of
the difficulties at various points throughout this paper.

\subsection{Technical Review of Fractional Superstrings and Parafermions}
\setcounter{footnote}{0}

We now provide a more technical review of
fractional superstrings and their constituent parafermion theories,
stressing only those aspects which will be necessary for later sections.
Complete details concerning the basic ideas behind fractional superstrings
can be found in Refs.~[1--4].

We begin by outlining some basic facts concerning the $\IZ_K$
parafermion theories \cite{Zamo,Gepner} which are ultimately the
building blocks of the fractional superstrings.
The $\IZ_K$ parafermion theory can be defined as the
coset theory $SU(2)_K/U(1)$ derived from the $SU(2)_K$
Wess-Zumino-Witten theory \cite{Knizhnik} after modding
out by a free $U(1)$ boson, but for our purposes we can simply think
of these $\IZ_K$ parafermion theories as a set of conformal field theories
with central charges
\beq
      c_\phi ~=~ {{2K-2}\over{K+2}}~,
\label{cphi}
\eeq
and with primary fields $\phi^j_m$ labeled and organized
by their $SU(2)$ quantum numbers $j$ and $m$, where $j\in \IZ/2$,
$j-m\in \IZ$, $|m|\leq j$,
and $0\leq j\leq K/2$.  These theories thus contain
only a finite number of fundamental fields, this number growing with
increasing $K$.  We shall generally define $\ell\equiv 2j$ and
$n\equiv 2m$, in terms of which the conformal dimensions (or
highest weights or spins) of the fields $\phi^j_m$
are given by
\beq
       h^{\ell}_{n}~\equiv~\Delta[\phi^j_m]~=~
      {{\ell(\ell+2)}\over{4(K+2)}} ~-~ {{n^2}\over{4K}} ~
\label{highestweights}
\eeq
and their fusion rules are given by
\beq
   \lbrack \phi^{j_1}_{m_1}\rbrack ~\otimes~
   \lbrack \phi^{j_2}_{m_2}\rbrack ~=~
   \sum_{J=|j_1-j_2|}^{J_{\rm max}} \,
   \lbrack \phi^{J}_{m_1+m_2}\rbrack ~
\label{parafusions}
\eeq
with $J_{\rm max} \equiv {\rm min}(j_1+j_2,K-j_1-j_2)$.
Here the sectors $[\phi^j_m]$ include the primary fields $\phi^j_m$
and their descendants.  The presence of the ``reflection'' symmetry
$j\leftrightarrow K/2-j$ allows us to consistently identify the
fields $\phi^j_m$ for values
outside the range $|m|\leq j$ via
\beq
       \phi^j_m~=~\phi^j_{m+K}~=~\phi^{K/2-j}_{-(K/2-m)}~.
\label{fieldidents}
\eeq
Note that according to (\ref{parafusions})
and (\ref{fieldidents}), those fields with $j,m\in\IZ$
constitute a closed subalgebra for even $K \geq 2$.

The complete worldsheet field content of the fractional superstring is then
taken to consist of a tensor product\footnote{
   This tensor-product formulation is to be distinguished from the
   ``chiral algebra'' formulation of fractional
   superstrings \cite{ALTtwo};  a summary of the possible relation between the
   two can be found in \cite{ALT} and \cite{AD}.}
of $D_c$ copies of this $\IZ_K$ parafermion conformal field theory,
tensored together with $D_c$
uncompactified (coordinate) bosons $X^\mu$.  Indeed, there exists a fractional
supersymmetry relating each coordinate boson to its corresponding $\IZ_K$
parafermion theory, with fractional supercurrent $J = \epsilon^\mu \partial
X_\mu + \eta$.  Here $\epsilon\equiv \phi^1_0$, and $\eta$ is defined to be
that parafermion descendent of $\epsilon$ which appears in the
$\epsilon^\mu (z)\epsilon_\mu(w)$ OPE with conformal dimension $1+h^2_0$.
The spacetime Lorentz indices $\mu =0,1,...,D_c-1$ are understood
to be contracted with the Minkowski metric.  The critical dimensions $D_c$ of
these theories depend on $K$, and can be determined
by a variety of arguments (see Refs.~\cite{AT,DT} for details) yielding
the result quoted in (\ref{kcritdimensions}).  Our cases of
interest are thus $K=2$, 4, 8, and 16 (yielding $D_c=10$, 6, 4, and 3
respectively).

With this tensor-product formulation,
it is easy to see that the fractional superstring reduces to the ordinary
superstring in the special case $K=2$.  The $\IZ_2$ ``parafermion''
theory has central charge $c_\phi=1/2$, and contains three independent fields
$\lbrace \phi^0_0,\phi^1_0,\phi^{1/2}_{1/2}\rbrace$ with conformal dimensions
$h=0,1/2,$ and $1/16$ respectively.  Thus the $K=2$ special case of the
$\IZ_K$ parafermion theory is readily identified as the Ising model, with the
above three ``parafermion'' fields identified respectively as the identity
$\bone$, the Majorana fermion field $\psi$, and the spin field $\sigma$.
(A fourth field $\phi^{1/2}_{-1/2}$ corresponds to the conjugate spin
field $\sigma^\dagger$.)  The worldsheet fractional supercurrent then reduces
to ordinary supercurrent $J=\psi^\mu \partial X_\mu$, since $\phi^1_0=\psi$
and no such parafermion field $\eta$ exists in the $\IZ_2$ theory.
The NS (spacetime bosonic) sectors of the superstring of course have
vacuum states corresponding to the primary fields $\phi^0_0=\bone$ or
$\phi^1_0=\psi$ in each direction, and the Ramond (spacetime fermionic)
sectors correspond only to the spin-field $\phi^{1/2}_{\pm 1/2} =\sigma$
in each direction.  Excitations in all sectors are effected via
$\phi^1_0=\psi$, which (according to the fusion rules) does not mix
NS or Ramond sectors with each other.

In analogous fashion, certain fields in the $\IZ_K$ parafermion
theories play special roles in the fractional superstring.
As is clear from the fusion rules (\ref{parafusions}),
the $\phi^0_0$ field continues to function as the identity for
all $K\geq 2$, whereas $\phi^{K/4}_{\pm K/4}$ fields
are the $K>2$ analogues of the Ising-model spin field $\sigma$.
The $\epsilon\equiv \phi^1_0$ field
is the analogue of the Ising-model field $\psi$,
and thus serves to generate Fock-space excitations in the
fractional superstring.
It is therefore possible to group many of the parafermion fields
into classes depending on the spacetime statistics
of the vacua they produce.  All of the fields which
close into each other under repeated
fusings of $\epsilon$ with itself and with the identity $\phi^0_0$
correspond to the various NS-like subsectors of the
theory, and all states in the towers built upon these vacua
are spacetime bosonic.  From the fusion rules (\ref{parafusions}),
we see that this set of NS-like primary fields
$\phi^j_m$ are those with $m=0$;  these were the fields collectively denoted
as $\lbrace\bone\rbrace$ in Sect.~2.1, producing
the various light-cone NS vacua $\lbrace\bone\rbrace^{D_c-2}$.
Similarly, the fields which close into each other
under repeated fusings of $\epsilon$ with the spin fields
$\phi^{K/4}_{\pm K/4}$ correspond to the Ramond-like subsectors
of the fractional superstring,
yielding towers of states which are spacetime fermionic.
These fields are of course those with $m=\pm K/4$,
denoted collectively as $\lbrace\sigma\rbrace$ in Sect.~2.1
and producing the various transverse Ramond-like vacua
$\lbrace\sigma\rbrace^{D_c-2}$.
Note that since the $m$-quantum numbers of these fields are additive
modulo $K/2$ under the fusion rules (\ref{parafusions}), this assignment
reproduces the expected spacetime boson/fermion selection rules:
\beq
      B \otimes B ~=~ B~,~~~~ B \otimes F ~=~ F~,~~~~ F \otimes F ~=~ B~.
\label{bosonfermionfusions}
\eeq
Obtaining (\ref{bosonfermionfusions}) is an important
consistency check on our identification of the spacetime statistics of
our parafermion states.

In order to most directly observe the presence of new sectors and
projections in these fractional-superstring theories for $K>2$,
let us construct their one-loop partition functions  $\calZ_K(\tau)$.
We shall concentrate on only those closed string theories in which
both the left- and right-moving theories exhibit a $K$-fractional
supersymmetry [this is the so-called $(K,K)$ fractional superstring,
the generalization of the ordinary Type-II superstring].
As usual, each worldsheet coordinate boson field $X^\mu$ contributes
to the partition function $Z(\tau)$ a factor
\beq
      \calZ_{\rm boson}~=~ {1\over{\sqrt{\tau_2} \,|\eta|^2}}~,
\label{bosonfactor}
\eeq
where $\eta(\tau)\equiv q^{1/24}\prod_{n=1}^\infty (1-q^n)$ is the
Dedekind $\eta$-function with $q\equiv \exp(2\pi i\tau)$ and
$\tau_2\equiv {\rm Im}\,\tau$;~
similarly, the contribution from each (chiral) parafermion field $\phi^j_m$
is given by \cite{Kac}
\beq
       (\calZ^j_m)_{\rm parafermion} ~=~  \eta \, c^{2j}_{2m}
\label{parafermionfactor}
\eeq
where the $c^\ell_n(\tau)$ are the
so-called ``string functions'' \cite{Distler}:
\beqn
    c^\ell_n(\tau) ~&=&~ q^{h^\ell_n + [4(K+2)]^{-1}} \,\eta^{-3}\,
      \sum_{r,s=0}^\infty (-1)^{r+s}\,q^{r(r+1)/2\,+
      s(s+1)/2\,+rs(K+1)}\,\times\nonumber\\
    &&~\times~\biggl\lbrace q^{r(j+m)\,+\, s(j-m)}~
      -~ q^{K+1-2j\,+\,r(K+1-j-m)\,+ \,s(K+1-j+m)}\biggr\rbrace
\label{stringfunctions}
\eeqn
and where the $h^\ell_n$ are the highest weights
given in (\ref{highestweights}).
We thus see that the string functions take the form
$ c^\ell_n = q^{H^\ell_n} (1+...)$ where only non-negative integer powers
of $q$ appear within the parentheses and where
\beq
         H^\ell_n~=~ h^\ell_n ~-~ {1\over 24} \left(1+c_\phi\right)~.
\label{Hvalues}
\eeq
The string functions exhibit the symmetries
\beq
    c^\ell_n ~=~ c^\ell_{-n}~=~ c^{K-\ell}_{K-n} ~=~ c^\ell_{n+2K}~,
\label{stringfunction_identities}
\eeq
and one conventionally defines the useful combinations
\beq
     d^\ell_n~\equiv~ c^\ell_n~+~ c^{K-\ell}_n
\label{ddefs}
\eeq
for $K\in 4\IZ$ and $\ell,n\in 2\IZ$.
These string functions $c^\ell_n$ can be viewed, of course, as
$K>2$ generalizations
of the Ising-model characters $\chi_0$, $\chi_{1/2}$, and $\chi_{1/16}$,
and indeed for $K=2$ we find
\beqn
       K=2:~~~~~~~~~~~\eta c^0_0 ~&=~ \chi_0~ &=~
           {1\over 2}\left( \sqrt{{{\vartheta_3}\over{\eta}}} +
            \sqrt{{{\vartheta_4}\over{\eta}}} \right)\nonumber\\
       \eta c^2_0 ~&=~ \chi_{1/2}~&=~
           {1\over 2}\left( \sqrt{{{\vartheta_3}\over{\eta}}} -
            \sqrt{{{\vartheta_4}\over{\eta}}} \right)\nonumber\\
       \eta c^1_1 ~&=~ \chi_{1/16}~&=~
       \sqrt{ {{\vartheta_2}\over{2\,\eta}}} ~.
\label{stringf_Ising}
\eeqn
We have written these Ising-model characters in terms of the
three Jacobi $\vartheta$-functions for later convenience.

 From (\ref{bosonfactor}) and (\ref{parafermionfactor}),
therefore, we expect the total partition function for the $(K,K)$
fractional superstring to take the form
\beq
   \calZ_K(\tau) ~=~ \tautwo^{1-D_c/2}\,
           \sum \,(\overline{c})^{D_c-2} (c)^{D_c-2}
             ~=~ \tautwo^{1-D_c/2}\,\sum_{m,n} \,a_{mn} \,\qbar^m q^n
\label{Zform}
\eeq
where the first summation is over all of the (as-yet undetermined)
contributing sectors of the theory, and where `$c$' schematically represents
the corresponding string-function factors.  The quantities $a_{mn}$ will
then represent the net number of states (spacetime bosonic minus fermionic)
with left- and right-energies $m$ and $n$ respectively.
In order to determine the contributing sectors, we simply
demand combinations of string functions in (\ref{Zform})
which not only render $\calZ_K(\tau)$ modular invariant,
but also yield $a_{nn}=0$ for all $n<0$ (absence of physical tachyons)
and $a_{00}\not= 0$ for at least {\it one}\/ sector (corresponding to
the existence of a graviton in the spacetime spectrum).
It turns out that this is sufficient\footnote{
    In the original derivations
    (see [1--3]), what is actually imposed is the stronger
    no-tachyon condition $a_{mn}=0$ if $m<0$ or $n<0$.
    This condition thus prevents the appearance of not only physical
    (on-shell) tachyons, but also unphysical (or off-shell) tachyons ---
    indeed, such a restriction is required for physically sensible theories
    of the Type-II [or $(K,K)$] variety.  It is this stronger condition
    which is ultimately responsible for both the vanishing and the uniqueness
    of the partition functions obtained.}
to yield the following set of {\it unique}\/
solutions [1--3] for each relevant value of $K\geq 2$:
\beqn
  \calZ_2&~=&~ \tautwo^{-4\phantom{/1}}\,\phantom{\bigl\lbrace}|A_2|^2
      \nonumber\\
  \calZ_4&~=&~ \tautwo^{-2\phantom{/1}}\,\left\lbrace|A_4|^2 ~+~ 3\,
   |B_4|^2\right\rbrace \nonumber\\
  \calZ_8&~=&~ \tautwo^{-1\phantom{/2}}\,\left\lbrace|A_8|^2 ~+~ |B_8|^2
     ~+~2\,|C_8|^2 \right\rbrace\nonumber\\
  \calZ_{16}&~=&~ \tautwo^{-1/2}\,\left\lbrace |A_{16}|^2 ~+~ |C_{16}|^2
   \right\rbrace
\label{partfuncts}
\eeqn
with
\beqn
  A_2&~=&~8\,(c^0_0)^7(c^2_0) +56\,(c^0_0)^5(c^2_0)^3
 +56\,(c^0_0)^3(c^2_0)^5 +8\,(c^0_0)(c^2_0)^7 -8\,(c^1_1)^8 \nonumber\\
  A_4 &~= &~4\,(c^0_0+c^4_0)^3(c^2_0)-4\,(c^2_0)^4-4\,(c^2_2)^4
   +32\,(c^2_2)(c^4_2)^3 \nonumber\\
  B_4 &~= &~16\, (c^0_0+c^4_0)(c^2_0)(c^4_2)^2
   +8\,(c^0_0+c^4_0)^2(c^2_2 c^4_2)-8\,(c^2_0)^2(c^2_2)^2 \nonumber\\
  A_8 &~= &~ 2\,(c^0_0+c^8_0)(c^2_0+c^6_0)-2\,(c^4_0)^2
   -2\,(c^4_4)^2+8\,(c^6_4c^8_4) \nonumber\\
  B_8 &~= &~ 4\,(c^0_0+c^8_0)(c^6_4)+4\,(c^2_0+c^6_0)(c^8_4)
   -4\,(c^4_0c^4_4) \nonumber\\
  C_8 &~= &~ 4\,(c^2_2+c^6_2)(c^0_2+c^8_2)-4\,(c^4_2)^2 \nonumber\\
    A_{16}&~= &~ c^2_0+c^{14}_0-c^8_0-c^8_8+2\,c^{14}_8 \nonumber\\
    C_{16}&~= &~ 2\,c^2_4+2\,c^{14}_4-2\,c^8_4~.
\label{ABC}
\eeqn
Note from (\ref{Hvalues}) that these combinations of string
functions each {\it a priori}\/ have $q$-expansions of the forms
\beq
     A_K\,\sim\, q^0\,(1+...)~,~~~
     B_K\,\sim\, q^{1/2}\,(1+...)~,~~~
     C_K \,\sim\, q^{3/4}\,(1+...)~,
\label{ABCexpansions}
\eeq
from which we deduce that only the $A$-type sectors contain massless
states.  Indeed, the string-function
combinations $+(D_c-2)(c^0_0)^{D_c-3} (c^2_0)$
and $-(D_c-2)(c^{K/2}_{K/2})^{D_c-2}$
within each of the $A_K$ combinations above
correspond to the (chiral) massless NS and Ramond vacuum states respectively;
note that they appear with the appropriate signs and normalizations for
bosonic spacetime vectors and fermionic spacetime spinors.
It is by tensoring the left- and right-moving (or holomorphic and
anti-holomorphic) sectors together that these chiral states produce
a massless $N=2$ supergravity multiplet.

For $K=2$, we can re-express $A_2$ in terms of the
equivalent $\vartheta$-functions to find that $A_2=\half \eta^{-12}J$
where $J\equiv \thetathree^4-\thetatwo^4-\thetafour^4$;
thus $\calZ_2$ is indeed recognized as the partition function of the ordinary
superstring, and the Jacobi identity $J=0$ (or $A_2=0$)
indicates the vanishing
of $\calZ_2$ (\ie, the exact level-by-level cancellation of bosonic
and fermionic states).  This is of course the partition-function
reflection of the spacetime supersymmetry of this theory.
Remarkably, however, this property extends to higher $K$ as well,
for it can be proven \cite{ADT} that {\it each}\/ of the combinations listed
in (\ref{ABC}) vanishes identically as a function of $q$:
\beq
     A_K~=~B_K~=~C_K~=~ 0 ~~~~~~~~~~~{\rm for~all~} K \geq 2~.
\label{ABCvanish}
\eeq
These resulting new identities, which are the higher-$K$ generalizations
of the $K=2$ Jacobi identity $A_2=0$,
can therefore be taken as evidence of spacetime
supersymmetry in the {\it fractional}\/ superstrings.
This is of course consistent with the appearance of the massless
supergravity multiplet in the fractional superstring spectrum, as noted
above.

We have now reached the point where the two fundamental issues
confronting us are evident.
Recalling (\ref{parafermionfactor}) and the discussion above
(\ref{bosonfermionfusions}), we see that it is a straightforward
matter to recognize whether each term within each $A_K$
arises from a sector which has a spacetime bosonic (NS)
vacuum state of the form
\beq
  {\rm NS}:~~~~~~~ \prod_{i=1}^{D_c-2} \phi^{j_i}_0~,
\label{NSvacua}
\eeq
or arises from a spacetime fermionic (Ramond) vacuum state of the form
\beq
   {\rm R}:~~~~~~~ \prod_{i=1}^{D_c-2} \phi^{j_i}_{\pm K/4}~.
\label{Rvacua}
\eeq
Thus, we can separate each of these expressions $A_K$ into
cancelling bosonic and fermionic pieces
\beq
           A_K ~=~ A_K^b ~-~ A_K^f
\label{Asplit}
\eeq
where
\beqn
   K=2:~~~~A_2^b ~&&\equiv
         ~8\,(c^0_0)^7(c^2_0) +56\,(c^0_0)^5(c^2_0)^3
         +56\,(c^0_0)^3(c^2_0)^5 +8\,(c^0_0)(c^2_0)^7  \nonumber\\
      A_2^f ~&&\equiv ~8\,(c^1_1)^8~  \nonumber\\
   K=4:~~~~A_4^b ~&&\equiv~ 2(d^0_0)^3 d^2_0-\quart(d^2_0)^4  \nonumber\\
      A_4^f ~&&\equiv ~\quart(d^2_2)^4-2(d^0_2)^3 d^2_2~  \nonumber\\
   K=8:~~~~
      A_8^b ~&&\equiv~ 2d^0_0 d^2_0-\half (d^4_0)^2  \nonumber\\
      A_8^f ~&&\equiv ~\half (d^4_4)^2-2d^0_4 d^2_4~  \nonumber\\
   K=16:~~~~
      A_{16}^b ~&&\equiv~ d^2_0-\half d^8_0  \nonumber\\
      A_{16}^f ~&&\equiv ~\half d^8_8-d^2_8~.
\label{Abf}
\eeqn
For $K>2$, we have made use of the identities
(\ref{stringfunction_identities})
and written $A_K^{b,f}$ in terms of
the $d^\ell_n$ combinations (\ref{ddefs}).

For the $B$- and $C$-type sectors, however, the situation is not
as clear.  It is readily seen from (\ref{ABC}) that
the $B$- and $C$-type terms arise from sectors with
vacuum states corresponding to primary-field combinations of the forms
\beqn
       B:&&~~~~~~~ \prod_{i=1}^{(D_c-2)/2} (\phi^{j_i}_0
     \phi^{j'_i}_{\pm K/4}) \nonumber\\
   C:&&~~~~~~~ \prod_{i=1}^{D_c-2}
        \phi^{j_i}_{\pm K/8}  ~,
\label{BCvacua}
\eeqn
and thus these sectors have vacua which do not even appear in the
ordinary superstring (indeed, no $B_2$ or $C_2$ terms appear
in the $K=2$ partition function).
These $B_K$ and $C_K$ terms do appear, however, in the partition
functions for $K>2$, forced upon us by modular invariance.
Although (\ref{ABCexpansions}) indicates that these sectors
contain only massive (\ie, Planck-scale) states,
arguments suggest \cite{ADT} that these sectors
play a crucial role in the ultimate consistency of the
fractional superstring.
It is therefore necessary, as a first step, to distinguish those
$B$- and $C$-sector states which are bosonic and fermionic --- \ie,
to achieve a splitting of the $B_K$ and $C_K$ partition-function
expressions into cancelling terms in analogy to that indicated in
(\ref{Asplit}) for the $A_K$-type sectors.
Recent results \cite{CR,AD} indicate, however,
that the appropriate $B_K$-sector splitting is as follows:
\beqn
   B_4^b ~&\equiv&~2\,d^0_0 d^2_2 d^0_0 d^4_2 - \quart\,
         (d^2_0d^2_2)^2 ~=~ 4\,q^{1/2}(1+...)\nonumber\\
   B_4^f ~&\equiv& ~ \quart\,(d^2_2d^2_0)^2 - 2\, d^4_2d^0_0 d^4_2 d^2_0
          ~=~ 4\,q^{1/2}(1+...)\nonumber\\
  B_8^b~&\equiv&~ 2\,d^0_0 d^6_4 - \half\,d^4_0 d^4_4
    ~=~ 2\,q^{1/2}(1+...) \nonumber\\
  B_8^f~&\equiv&~ \half\,d^4_4 d^4_0 - 2\,d^8_4 d^2_0 ~=~ 2\,q^{1/2}(1+...)~,
\label{Bbf}
\eeqn
whereas the corresponding $C_K$-sector splitting indicates that these sectors
contain no physical states whatsoever:
\beqn
  C_8^b~&\equiv&~ 2\,d^0_2 d^2_2 - \half\,(d^4_2)^2
    ~=~ 0 \nonumber\\
  C_8^f~&\equiv&~ \half\,(d^4_2)^2 - 2\,d^0_2 d^2_2
    ~=~ 0 \nonumber\\
  C_{16}^b~&\equiv&~ d^2_4 - \half\,d^8_4
    ~=~ 0 \nonumber\\
  C_{16}^f~&\equiv&~ \half\,d^8_4 - d^2_4
    ~=~ 0~.
\label{Cbf}
\eeqn
This absence of states will be discussed below and in Sect.~5.3.
In Ref.~\cite{AD} a proposal was made whereby the $B_K$-sector
splitting in (\ref{Bbf}) could be
understood in terms of the {\it ordering}\/ of the parafermion
$m$-quantum numbers --- {\it e.g.},
\beqn
     {\rm bosonic} ~\Longleftrightarrow~
        &&m_i~=~\left( 0,K/4,0,K/4,...,0 , K/4\right)\nonumber\\
     {\rm fermionic} ~\Longleftrightarrow~
        &&m_i~=~\left( K/4,0,K/4,0,...,K/4,  0\right)
\label{assignment}
\eeqn
where each factor (0 or $K/4$) is repeated $(D_c-2)/2=8/K$ times.
Indeed, this assignment reproduces the proper statistics selection
rules (\ref{bosonfermionfusions}) not only within the
$B$-sectors, but also between the $A$- and $B$-sectors.
The necessity for such an ordered assignment, however, illustrates
that the $B$-sectors appear to break permutation invariance amongst
the $D_c-2$ different copies of the fundamental boson plus $\IZ_K$ parafermion
theory.  This loss of permutation invariance is thus inconsistent
with the original formulation of the fractional superstring
in which the worldsheet conformal field theory is taken to be a tensor
product.

This problem is connected with the second fundamental issue,
the ``internal projections''.
The expressions in (\ref{Abf}) are written in terms of
the string functions $c^\ell_n$ and $d^\ell_n$, each of which
is the character of a particular
sector of the joined boson plus parafermion system.  Hence these
expressions serve as a means of tallying the degeneracies $g_n^{b,f}$ of
spacetime bosonic or fermionic states
at each mass level $n$ in the parafermion plus boson Fock space:
\beq
            A_K^{b,f}~=~ \sum_{n=0}^\infty \,g_n^{b,f}\,q^n~.
\label{Abfexpand}
\eeq
As such, it is easy to interpret the expressions $A_2^b$ and $A_2^f$
for the $K=2$ superstring case:  the summations within each of these separate
expressions in (\ref{Abf}) simply represent the added contributions from each
of
the bosonic or fermionic sectors in the theory.  For $K>2$, however,
such an interpretation for each $A_K^b$ and $A_K^f$ becomes more difficult,
since the contributions of certain parafermionic sectors appear to
be {\it subtracted}\/ rather than added.
We shall refer to those parafermionic sectors whose contributions are
subtracted as ``internal projection sectors'';
note that their presence is, like
that of the $B$- and $C$-sectors, forced upon us by modular invariance.
It is straightforward to verify that
these internal projection sectors themselves contain only massive (\ie,
Planck-scale) states, so the subtractions they introduce into
the net state degeneracies $g_n^{b,f}$ in (\ref{Abfexpand}) appear only
for $n\geq 1$.  Thus the states at the {\it massless}\/ levels of
the bosonic and fermionic $A$-sectors
(including the supergravity multiplet) are unaffected.

The effects of these internal projections are nevertheless quite
profound.  The GSO projection, for example, is what has
prevented certain sectors from ever appearing within our
partition function expressions, so we may say that effectively
an entire tower of states has been projected out of the spectrum.
These internal projections, on the
other hand, seem to subtract one sector from a {\it different}\/ one,
leaving behind a set of state degeneracies $g_n^{b,f}$
in (\ref{Abfexpand})
which may or may not be physically sensible from a worldsheet conformal field
theory point of view.
It can easily be verified that despite the internal projection,
the values of each $g_n^{b,f}$ within (\ref{Abfexpand})
are all non-negative, so it remains to discover whether
the expressions $A_K^{b,f}$ can themselves be consistently
understood as the characters
(or even the {\it sum} of characters)
of the highest-weight sectors of some new conformal field theory:
\beq
          A_K^{b,f} ~{\buildrel ? \over =} ~ \sum_h\,\chi_{\rm h}~.
\label{maybecharacter}
\eeq
It turns out, however, that the
expressions $A_K^{b,f}$ all pass this first non-trivial test,
for it can be shown \cite{ADT} that
\beq
          A_K^{b,f}(\tau) ~=~
     (D_c-2)\, \left({1\over\eta}\right)^{D_c-2}\,
      \left[\chi_{1/16} \right]^{D_c-2} ~=~
            (D_c -2) \, \left[
      {{\vartheta_2(\tau)}\over{2\,\eta^3(\tau)}} \right]^{(D_c-2)/2}
\label{Aintproj}
\eeq
where $\chi_{1/16}$ is the Ising-model character given in
(\ref{stringf_Ising}).
For $K=2$, of course, this relation is manifestly true as a
consequence of (\ref{stringf_Ising}).  For $K>2$,
however, this is a truly remarkable result, indicating that
despite our original conformal-field-theoretic formulation
in terms of bosons and parafermions,
the numbers of states surviving the internal projections
at each mass level of the theory are precisely those
of $D_c-2$ copies of a single
boson plus {\it fermion}, or $(X,\psi)$, theory!
This in turn implies that exponentially large numbers of
states are being projected out of the spectrum by these new
internal projections for $K>2$, for while the original boson/parafermion
theory in light-cone gauge has total central charge
\beq
      c~=~(D_c-2)\,(1+c_\phi)~=~{{48}\over{K+2}}~,
\label{precentralcharge}
\eeq
the post-projection theory in light-cone gauge has the smaller central charge
\beq
      c'~=~(D_c-2)\,(1+ c_\psi)~=~{{24}\over K}~.
\label{postcentralcharge}
\eeq
Indeed, only for $K=2$ are these two central charges equal.

A similar situation exists for the $B$-sectors, for it is readily
seen from (\ref{Bbf}) that analogous internal projections appear within the
expressions $B_K^{b,f}$ for both $K=4$ and $K=8$.
The coefficients within a $q$-expansion of these expressions are again
all positive, however, and remarkably there again exists a
simple identity \cite{AD} analogous to (\ref{Aintproj}):
\beq
          B_K^b(\tau)~=~ B_K^f(\tau) ~=~ (D_c -2) \, \left[
      {{\vartheta_2(\lambda\tau)}\over{2\,\eta^3(\tau)}}
       \right]^{(D_c-2)/2}~
\label{Bsplit}
\eeq
where
\beq
        \lambda ~\equiv ~ (\Delta[\epsilon])^{-1} \,=\, (h^2_0)^{-1}\,=\,
    \half\,(K+2) ~=~ \cases{  3 & for $K=4$ \cr 5 & for $K=8$~.\cr}
\label{lambdadef}
\eeq
Thus, whereas $A_K^{b,f}$ are related to the Dedekind $\eta$-function
and Jacobi $\vartheta$-functions whose arguments were the usual torus
modular parameter $\tau$, we find that our $B$-sector characters $B_K^{b,f}$
are given by similar expressions in which the arguments of the
$\vartheta$-functions are now rescaled by factors of $\lambda$,
where $\lambda$ is in general the inverse spin of the
parafermion field $\epsilon\equiv \phi^1_0$ in the $\IZ_K$ parafermion
theory.  As we shall see in Sect.~4, Eq.~(\ref{Bsplit}) indicates that
the numbers of states surviving the internal projections
at each mass level of the $B$-sectors of the theory are precisely those
of $D_c-2$ copies of a single
boson plus fermion theory, where this ``$B$-sector fermion'' is
now formulated on a {\it rescaled}\/ internal momentum lattice.
This momentum-lattice rescaling for the fermion does not alter
its central charge contribution, however, and thus
(\ref{Bsplit}) also implies the post-projection central-charge reduction
(\ref{postcentralcharge}).
This is of course necessary for the consistency of the internal
projections in both the $A$- and $B$-sectors.
We remark in passing that it is gratifying to observe the
original parafermion spin re-emerging in so non-trivial a manner, for
the internal projections in $A$-sector seemed, by projecting each
copy of our fundamental parafermion theory down to the Ising model
({\it i.e.}, to a free fermion), to have erased all knowledge of the
fractional spins which were the original
starting point in the fractional-superstring
construction.  We also note that these internal projections
provide a natural explanation for the absence of physical
states in the fractional-superstring $C$-sectors (\ref{Cbf}):
these states are not GSO-projected, but rather {\it internally projected}\/
out of the spectrum.   This will be discussed in Sect.~5.3.

Despite the appearances of (\ref{Aintproj}) and (\ref{Bsplit}),
we cannot simply declare that the post-projection conformal field theory for
the $A_K$ and $B_K$ sectors is that of $D_c-2$ coordinate bosons tensored
together with $D_c-2$ independent copies of the Ising model (or the
``rescaled'' Ising model);  as we will see, the conformal field theory this
would produce turns out to be far too large to correctly describe the
fractional-superstring spectrum of states.  Furthermore, as our above
arguments suggest, this smaller post-projection CFT is not expected to
have a simple tensor-product formulation.  Therefore, in order to determine
the correct post-projection conformal field theory for the $A$- and
$B$-sectors,
we must carefully map out the parafermionic sectors which survive the internal
projections, and attempt to describe them in terms of the Fock space of some
new Ising-like conformal theory with central charge (\ref{postcentralcharge}).
It is towards this endeavor that we now turn.


\section{Internal Projections in the $A$-Sector}
\setcounter{footnote}{0}

In this section we explore the conformal field theory (CFT) describing
the $A$-sector of the fractional superstring in light-cone gauge
after the internal projection.
We will find that we can determine its spectrum of highest weights
as well as its fusion rules, and although we cannot presently
construct an adequate representation of this conformal field
theory in terms of worldsheet fields,
our analysis will provide a minimal set of constraints which this
representation must satisfy and which can hopefully serve as a guide towards
its ultimate identification.
Our starting point will be Eq.~(\ref{Aintproj}), and indeed by
using this relation we have already seen
that the central charge of this CFT must be given by
(\ref{postcentralcharge}).
In this section we will more fully and systematically
examine the consequences of the relation (\ref{Aintproj}).

Our fundamental approach --- and indeed our philosophy
concerning these internal projections --- can be described as follows.
Although the {\it pre}-projection conformal field theory was
uniquely identified (by construction) as a tensor product of $D_c-2$
free bosons and $D_c-2$ copies of the $\IZ_K$ parafermion theory,
the resulting internal projections indicate that the Fock
space of this original CFT is far too large for describing what
we know (via the ``top-down'' partition function analysis)
to ultimately be the physical states in the fractional-superstring spectrum.
Therefore, while our present description of the fractional superstring
involves this large parafermionic worldsheet CFT in conjunction with
an as-yet mysterious internal projection, it is
expected (and in fact required for self-consistency)
that there exist an alternative
description starting directly with the smaller post-projection CFT
of central charge (\ref{postcentralcharge}) {\it in which no
internal projection appears}.
Such a description would clearly be preferable,
for it is not presently known how the internal projection is to be
implemented in terms of the original parafermion fields \cite{ALT}.

We begin by focusing on the spectrum of states of this
post-projection CFT, for obtaining a clear description
of its spectrum is a necessary prerequisite for its complete
identification.  Since this smaller CFT is in
some sense {\it embedded}\/ within the original larger parafermionic
CFT, we expect its Fock space to be describable in two ways:
as {\it all}\/ of the states of the smaller Ising-like CFT,
and as only certain selected states in the large parafermion
CFT.  In Sect.~3.1, therefore, we shall first introduce the Ising-like
CFT which turns out (along with the $D_c-2$ coordinate bosons)
to be the post-projection
CFT for the $A$-sector.  In Sect.~3.2 we shall then explicitly
determine which of the sectors in the parafermion theory are those
which ultimately survive this internal projection and yield
this Ising-like CFT.

\subsection{The Post-Projection CFT and the Ising Model}

Our fundamental starting point, and indeed the sole
indication of any relation between the $A$-sector post-projection CFT and
the Ising model, comes from (\ref{Aintproj}):  this equation
indicates that the post-projection CFT in the $A$-sector
contains the Ramond spin state $\sigma^{D_c-2}$ whose character
is $(\chi_{1/16})^{D_c-2}$.  Since the $\sigma$ field is
a primary field of the Ising model, it is straightforward
to demonstrate that any such CFT containing
the state $\sigma^{D_c-2}$ must also
contain the identity ``vacuum state'' $\bone^{D_c-2}$
as well as the fermion-field state $\psi^{D_c-2}$.
The first complication that we face, however,
is that there in general exist {\it many}\/ different self-consistent
``Ising-like'' conformal field theories which contain all three
of these states.

To demonstrate this, let us consider the situations for different
values of $D_c$.
In the $K=16$ case, we have $D_c-2=1$, and
thus it is of course clear that our light-cone ``Ising-like'' CFT
can be nothing but the
Ising model itself, with three primary fields $\lbrace
\bone,\psi,\sigma\rbrace$
with highest weights $h=\lbrace 0,1/2,1/16\rbrace$ respectively,
and with fusion rules of the general form
\beqn
    \lbrack\varphi_1\rbrack \times [\varphi_1] ~&=&~ [\bone] \nonumber\\
    \lbrack\varphi_1\rbrack \times [\varphi_2] ~&=&~ [\varphi_2] \nonumber\\
    \lbrack\varphi_2\rbrack \times [\varphi_2] ~&=&~ [\bone] + [\varphi_1]
\label{Isingfusions}
\eeqn
where we identify $\lbrace\bone,\varphi_1,\varphi_2\rbrace
\leftrightarrow \lbrace \bone,\psi,\sigma\rbrace$.
For the $K=8$ case, however, we have $D_c-2=2$, and
it turns out that there are {\it two}\/ possible self-consistent
CFT's which incorporate the three states $\bone^2$, $\psi^2$, and $\sigma^2$.
The first such theory is a simple tensor product of the two Ising models:  this
tensor-product (Ising)$^2$ theory has $c=1$ and
 {\it a priori}\/ contains a total of $3\times3=9$ ``primary
fields''.\footnote{
    For $c\geq 1$ CFT's, there are of course an
    infinite number of fields which are primary with respect to the Virasoro
    algebra. We are here adopting a somewhat looser terminology when speaking
    of such $c\geq 1$ theories, grouping these primary fields according
    to their highest weights (mod 1), and referring to this collected
    set of primary fields (as well as all of their Virasoro descendants) as
    a single ``sector'' as though it were generated by a single primary
    field.   The associated fusion rules and characters must
    then be understood in this context.}
The second such theory, however, is the so-called $c=1$ Dirac-fermion CFT:
this theory is a subset of (Ising)$^2$,
and represents the ``diagonal'' combination of two
Ising models (as is particularly
easy to see in a toroidal boundary-condition basis for the two Ising-model
Majorana fermions).
This Dirac-fermion CFT is equivalent to a free boson
compactified on a circle of radius $R=1$, and
contains only the three primary fields corresponding to
the three winding-mode sectors which exist at this radius.
These primary fields, which we can denote $\bone$,
$\varphi_1$, and $\varphi_2$, have highest weights
$h=0,1/2,$ and $1/8$ respectively, and this theory turns out to
have the same fusion rules (\ref{Isingfusions}) as the Ising model
itself.  The relations between the three characters of
the Ising model given in (\ref{stringf_Ising}), and the three characters
$\chi_h^{(c=1)}$ of this Dirac-fermion CFT, are as follows:
\beqn
      \chi_0^{(c=1)} ~&=&~ (\chi_0)^2 + (\chi_{1/2})^2 \nonumber\\
      \chi_{1/2}^{(c=1)} ~&=&~ \chi_0\,\chi_{1/2} \nonumber\\
      \chi_{1/8}^{(c=1)} ~&=&~ (\chi_{1/16})^2~.
\label{chars_one}
\eeqn
 From these relations it is thus easy to see that
the $\sigma^2$ state is contained within the $h=1/4$
sector of the Dirac fermion theory, while the $\bone^2$ and $\psi^2$
states are contained within the $h=0$ (vacuum sector) of this theory.

A similar situation exists for the $K=4$ case with $D_c-2=4$.
Here again there are several
different Ising-like CFT's containing the three states $\bone^4$, $\psi^4$,
and $\sigma^4$;  examples include a tensor product of four $c=1/2$
Ising models, or a tensor product of two $c=1$ Dirac fermion theories.
There exists again, however,
an even smaller $c=2$ diagonal combination of the two Dirac fermion
theories which, like the $c=1$ Dirac theory above,
contains only three primary field sectors.
These ``fields'' have highest weights $h=\lbrace 0,1/2,1/4\rbrace$
respectively, and this theory too has the fusion rules (\ref{Isingfusions}).
The characters corresponding to these three primary fields are given by
\beqn
      \chi_0^{(c=2)} ~&=&~
    \left\lbrack \chi_0^{(c=1)}\right\rbrack^2 +
       \left\lbrack \chi_{1/2}^{(c=1)}\right\rbrack^2
 ~=~ \half\,\left\lbrack(\chi_0+\chi_{1/2})^4
       + (\chi_0-\chi_{1/2})^4\right\rbrack \nonumber\\
      \chi_{1/2}^{(c=2)} ~&=&~
   \chi_0^{(c=1)} \chi_{1/2}^{(c=1)}
  ~=~\textstyle{1\over 8}\,
        \left\lbrack(\chi_0+\chi_{1/2})^4
       - (\chi_0-\chi_{1/2})^4\right\rbrack \nonumber\\
      \chi_{1/4}^{(c=2)} ~&=&~
    \left\lbrack \chi_{1/8}^{(c=1)}\right\rbrack^2 ~=~ (\chi_{1/16})^4~.
\label{chars_two}
\eeqn
Once again we see that the $\sigma^4$ state is contained within the
$h=1/4$ sector of this theory, while the $\bone^4$ and $\psi^4$ states are
contained within the vacuum sector with $h=0$.

Finally, even in the $K=2$ superstring case, there are {\it a priori}\/
a variety of choices.
The largest is of course a tensor product of eight Ising models,
and the smallest, analogously, is again a certain $c=4$ diagonal combination
of the two $c=2$ diagonal theories above.  This latter theory
again contains three primary field sectors,
one with highest weight $h=0$ and two with $h=1/2$,
and has the same fusion rules as the Ising model itself.
Its three characters are given by
\beqn
      \chi_0^{(c=4)} ~&=&~ \half\,\left\lbrack(\chi_0+\chi_{1/2})^8
       + (\chi_0-\chi_{1/2})^8\right\rbrack \nonumber\\
       \chi_{1/2}^{(c=4)} ~&=&~ \textstyle{1\over {16}}\,
        \left\lbrack(\chi_0+\chi_{1/2})^8
       - (\chi_0-\chi_{1/2})^8\right\rbrack \nonumber\\
      \tilde {\chi}_{1/2}^{(c=4)} ~&=&~ (\chi_{1/16})^8~.
\label{chars_four}
\eeqn
Note that in this case  we now have {\it two} characters
with $h=1/2$;  these characters are in fact equal as functions of $q$
as a result of the Jacobi identity
$\chi_{1/2}^{(c=4)}=\tilde \chi_{1/2}^{(c=4)}$.

Choosing the appropriate Ising-like CFT from among these various
possibilities for each value of $K$ is of paramount importance,
for while they all share the same ground state for each value of $K$,
they each have drastically different sets of primary fields
and fusion rules.
Our guide shall be the $K=2$ case, for in the ordinary $D_c=10$ superstring
we know precisely which Ising-like CFT is ultimately the one required for
self-consistency:  it is the smallest or ``minimal''
CFT mentioned above, containing only three sectors.
In the usual parlance, the sector corresponding
to $\tilde \chi_{1/2}^{(c=4)}$ is of course the Ramond sector $\sigma^8$,
while the two sectors corresponding to
$\chi_0^{(c=4)}$ and $ \chi_{1/2}^{(c=4)}$ are the NS$^{(\pm)}$
sectors of the superstring with odd and even $G$-parities respectively.
The NS$^{(-)}$ sector with odd $G$-parity of course contains spacetime
tachyons (since it has $h=0$), and it is
removed by the GSO-projection.
Thus, from (\ref{chars_four}), we see that
the subsectors $\bone^i\psi^{D_c-2-i}$ with even $i$
are projected out of the spectrum (as we discussed in
Sect.~2.1), while those with odd $i$ remain.
That the Ramond and NS$^{(+)}$ sectors are
are the only two surviving sectors of the superstring
can be easily seen from the superstring partition function, which
takes the simple form:
\beq
    \calZ_{K=2}~=~ 64~\tautwo^{-4} ~ |\eta|^{-16}~
      \left| \, \chi_{1/2}^{(c=4)} \,-\,
        \tilde \chi_{1/2}^{(c=4)} \,\right|^2~=~0~.
\label{superpartf}
\eeq
Thus, while the ordinary superstring could equivalently be described in terms
of any of the larger Ising-like $c=4$ theories considered above, one would
have to assume all of the extraneous sectors thus introduced to be
``GSO-projected'' out of the spectrum so that (\ref{superpartf})
and the corresponding minimal $c=4$ theory are ultimately obtained.
This $c=4$ theory is of course nothing but the
$SO(8)_1$ Wess-Zumino-Witten (WZW) model, with all states
forming representations of the transverse Lorentz group
$SO(8)$.

The fact that there are no mixed Ramond/NS sectors
in the superstring means, as we have already noted,
that it is straightforward to identify the spacetime spin-statistics
of all of the states in the Fock space of the superstring,
and indeed all states transform under $SO(8)$ with either integer
or half-integer spins.
As we have seen in Sect.~2.3, however, the analogous situation
exists in the $A$-sectors of the fractional superstring:  all
of the states in the $A$-sectors arise from vacuum states of the NS
or Ramond types given in (\ref{NSvacua}) and (\ref{Rvacua}) respectively,
and have well-determined spacetime spin-statistics.
Therefore, we shall assume that the appropriate $A$-sector light-cone
CFT's implied by (\ref{Aintproj}) for each value of $K$ are
the ``minimal'' or diagonal Ising-model combinations discussed above:
\beq
   K\geq 2:~~~~~~
  \left(c'={24\over K}\right)~{\rm CFT} ~=~
     \left\lbrace \mathop\otimes_{\mu=1}^{D_c-2=16/K}  X^\mu \right\rbrace
          ~~\otimes~~
  \left\lbrace \left( c={8\over K}\right)~{\rm minimal~theory}\right\rbrace~;
\label{CFTchoice}
\eeq
indeed, these ``minimal'' Ising-like CFT's are equivalent to
$SO(D_c-2)_1$ WZW models.
These choices (\ref{CFTchoice}) are thus not only free of the
mixed Ramond/NS sectors and in agreement with the ordinary superstring
for the $K=2$ special case,
but are also (as we shall find below) the only ones consistent with
the parafermionic CFT's and their internal projections.
We caution, however, that (\ref{CFTchoice}) does not indicate
how this CFT is to ultimately be {\it represented}\/ in terms of
an appropriately chosen set of worldsheet fields so
that a spacetime string theory with the correct spacetime
statistics properties might be constructed.
These issues will be discussed in Sects.~3.3~and~6.

\subsection{The Post-Projection CFT in Parafermion Language}

Given the post-projection CFT in (\ref{CFTchoice}),
we now seek to understand how it arises as the
result of internal projections between certain parafermion sectors
in the original $c=48/(K+2)$ worldsheet theory.

As a consequence of this decreasing of the central charge
induced by the internal projections,
not all of the parafermion sectors which
have played a role in the fractional superstring
prior to the internal projection can be expected to survive to play
an analogous role in the residual post-projection CFT.
To illustrate this point, let us consider
the lowest parafermion vacuum state in the $K=4$ theory.
This state $(\phi^0_0)^4$ is tachyonic, with $({\rm mass})^2= -1/3$,
and although this state (along with the entire Fock space of states built upon
it) ultimately suffers a GSO-projection and does not appear
in the modular-invariant partition function (\ref{partfuncts}),
this state still serves as the ultimate {\it ground state} of the
parafermionic worldsheet theory.  More precisely, in
conformal-field-theoretic language, this field $(\phi^0_0)^4$
serves as the identity field in the $c=48/(K+2)=8$ parafermionic
tensor-product theory, and has highest weight $h= 4 h^0_0 =0$.  From
(\ref{Aintproj}), however, we have determined that the
post-projection CFT has central charge $c'=24/K=6$,
and therefore the ground state of the post-projection
CFT should only have $({\rm mass})^2 = -c'/24= -1/K=-1/4$.
This implies that the state $(\phi^0_0)^4$, although the true ground
state of the {\it parafermionic}\/ tensor-product CFT, is
clearly {\it not}\/ the ground state of the smaller post-projection
Ising-like theory --- indeed, it is effectively
not even {\it in}\/ this post-projection theory at all.
Similar conclusions hold for the $K=8$ and $K=16$ cases as well.

Which parafermion state actually serves, then, as the true ground
state of the post-projection CFT for each general value of $K$?
We are of course guaranteed that such a state exist, for all
states in the Ising-like post-projection CFT must also arise as states
in the original parafermionic CFT.
In the language of the minimal Ising-like model with $c=8/K$,
this ground state is of course easily identified as
the state $\bone^{D_c-2}$ within the $h=0$ sector,
but in the language of the larger parafermion
theory with central charge $c=48/(K+2)$, this ground state must appear
as some excited state with highest weight $H_0$ and $({\rm mass})^2=-1/K$.
[Indeed, only the quantity (mass)$^2$ is ``invariant'' under
change of the CFT being used to describe the spectrum.]
Thus, since in general $({\rm mass})^2=H-c/24$, we obtain
\beq
       H_0 ~=~ {c\over{24}} \,-\,{1\over K}~=~
          {2\over {K+2}} \,- \,{1\over K}~,
\label{Hground}
\eeq
and for each $K$ this can be rewritten in terms of the highest
weights $h^\ell_n$ of the parafermion theory as
\beq
       H_0 ~=~ (D_c-3)\,h^0_0 \,+  \,h^2_2~.
\label{Hgroundtwo}
\eeq
Thus we see that for each value of $K$, it is the state
\beq
        {\rm effective~vacuum:} ~~~~~~ (\phi^0_0)^{D_c-3} \,\phi^1_{\pm 1}~,
      ~~~ ({\rm mass})^2 ~=~ -1/K
\label{effectivevacuum}
\eeq
which serves as the effective vacuum of the post-projection CFT
in the $A$-sector.  Note that $h^\ell_n=h^\ell_{-n}$ even though
$\phi^j_m \not= \phi^j_{-m}$:  it is for this reason that there
exists an ambiguity in identifying the sign of the $m$-quantum numbers
of parafermion fields in this approach.
Either choice of sign, however, yields a state at
the same $({\rm mass})^2$ level.
We also note, of course, that only for the $K=2$ superstring case is
this effective vacuum state in (\ref{effectivevacuum}) equivalent
to the original state
$(\phi^0_0)^{D_c-2}$;  this follows as a consequence of (\ref{fieldidents}).
For other values of $K$,
it is the inequality of these two states which reflects
the reduction in the central charge induced by the internal projection.
Indeed, as a general rule, the less negative the (mass)$^2$ of the
ground state of a conformal field theory, the smaller its central charge.

One point deserves special emphasis:
the fact that (\ref{effectivevacuum}) serves as
the effective vacuum of the post-projection CFT in the $A$-sector
does not imply that there exist spacetime particles of $({\rm mass})^2=
-1/K +n$, $n\in \IZ$ in the fractional-superstring spectrum.
Indeed, we know from the partition functions (\ref{partfuncts})
that no such states exist in the $A$-sectors either before or after
the internal projections:  the $A$-sectors contain only states with
$m^2\in \IZ$.  The state (\ref{effectivevacuum}), along with
the entire tower of states built from it, therefore experiences a GSO
projection and fails to appear in the spectrum in the same way the original
vacuum state $(\phi^0_0)^{D_c-2}$ failed to appear;  it is only the
effect of the internal projection that the latter vacuum state has been
replaced by the former.

Although the state in (\ref{effectivevacuum}) serves as the
effective vacuum of the internally projected $A$-sector,
the entire parafermionic sector built upon this vacuum state
cannot by itself comprise the corresponding sector of the internally projected
theory, for we still must incorporate the internal projection.
Specifically, this means that there must be at least one other
sector in the parafermion theory whose Fock space of states
must be {\it subtracted}\/
or removed from those in the above vacuum-state sector
in order to produce a residual Fock space appropriate for a CFT with $c=24/K$.
Such a subtraction would be analogous to those appearing in
(\ref{Abf}) for $K>2$.

It is a simple matter to determine these various potential
``projection sectors''.
Since the states from two sectors can be subtracted from each
other in this way only if they share the same (mass)$^2$ values, any such
possible other ``projection sectors'' would have to have vacuum states with
highest weights $H'_0$ differing from $H_0$ by integers.
For $H'_0=H_0+1$, we find that there exist in general {\it two}
such possible states, for this value of $H'_0$ can be written in terms of the
parafermionic highest weights in two distinct ways:
\beq
       H'_0 ~=~  (D_c-3)\,h^{K/2}_0 + h^{K/2}_2 ~=~
       (D_c-4)\,h^0_0 +  h^2_0 + h^K_{K-2}~.
\label{potproj}
\eeq
[The second expression in (\ref{potproj}) is of course appropriate only for
the cases in which $D_c\geq 4$ --- {\it i.e.}, for $K=4$ and $K=8$.]
We thus have, at this mass level, the two possible projection sectors
\beq
       ({\rm mass})^2 = 1-{1\over K}:~~~~~~~~
    (\phi^{\pm K/4}_0)^{D_c-3} \phi^{K/4}_{\pm 1} ~,~~~
    (\phi^0_0)^{D_c-4}\phi^1_0 \,\phi^{K/2}_{\pm(K/2-1)}~.
\label{potprojtwo}
\eeq
It turns out that we need not consider any higher values of $H'_0$,
however, for all other solutions can be generated from those above.
Since $h^{K-\ell}_n - h^\ell_n = \half(K/2-\ell)$,
we see that $\phi^j_m$ and $\phi^{K/2-j}_m$ have highest
weights differing by integers when $K\in 4\IZ$ and $j\equiv \ell/2\in \IZ$.
Thus, for each of the potential projection sectors listed above,
there are others in which each $\phi^j_m$ is replaced by $\phi^{K/2-j}_m$,
and these variations indeed fill out the entire space of
solutions for all values
of $H'_0$.  We shall not explicitly list these other options, but they will
be included in what follows.  Thus, we expect
that for each value of $K$, some linear combination of
all of these sectors will yield the $h=0$ sectors of our minimal
Ising-like post-projection CFT's;
the presence of negative coefficients in these linear combinations
will then indicate those sectors serving as projection sectors.

It is a relatively straightforward matter to determine these linear
combinations, for the Ising-like characters $\chi^{(c=8/K)}_0$ of these
$h=0$ sectors have already been determined in (\ref{chars_two})
and (\ref{chars_four}), and the
chiral character of $16/K$ free uncompactified bosons is
simply $\eta^{-16/K}$.
Thus, we seek a linear combination of the characters corresponding
to each of the above potential sectors which reproduces these
Ising-like characters.
Remarkably, such a combination exists for each value of $K$.
Let us define the three quantities:
\beqn
    K=4:&~~~~~~~~~ U_4 ~&\equiv~\half (d^0_0)^3 d^2_2 ~+~
        {\textstyle{3\over 2}} (d^0_0)^2 d^2_0 d^4_2
         ~-~{\textstyle{1\over 4}} (d^2_0)^3 d^2_2 \nonumber\\
    K=8:&~~~~~~~~~ U_8 ~&\equiv~ d^0_0 d^2_2 ~+~
           d^2_0 d^8_2
         ~-~ \half d^4_0 d^4_2 \nonumber\\
    K=16:&~~~~~~~~~U_{16}~&\equiv~   d^2_2 ~-~ \half  d^8_2~,
\label{Avacids}
\eeqn
where the $d^\ell_n$ functions were defined in (\ref{ddefs}).
Then we indeed find
\beq
         U_K ~=~ \eta^{-16/K}\,\chi_0^{(c=8/K)}
\label{Avacidstwo}
\eeq
where the $\chi_h^{(c=1/2)}$ characters are of course those
of the Ising model.
Thus, we see that it is always the {\it first}\/ of the sectors listed
in (\ref{potprojtwo}) which serves as the projection
sector in each of the relevant cases, whereas the second sector
(when present) merely contributes to the Fock space {\it prior}\/ to
the internal projection.
Note that while each of the linear combinations on the left sides
of the above equation involves
a subtraction, as required for central charge reduction,
no minus signs ever appear within the expressions
on the right sides.
Also note that the coefficients in (\ref{Avacids}) are
always integers when these equations are
expressed directly in terms of the individual string functions
$c^\ell_n$ (rather than the combinations $d^\ell_n$).  Thus, we indeed have
a proper mapping between the internally projected parafermion theory
and the $h=0$ sector of the minimal Ising-like theory for each value
of $K$.
Furthermore, just as in the $K=2$ superstring case, this $h=0$ sector
is GSO-projected out of the fractional superstring spectrum.

Having thus isolated the $h=0$ ground state of the post-projection CFT,
we now proceed to the next-lowest state.
The next-to-lowest vacuum state in the minimal Ising-like CFT
for $K>2$ is clearly that with $h=1/K$;  note that this state
$\sigma^{D_c-2}$ thus sits higher than any of
the $\bone^{D_c-2-i} \sigma^i$ states with $h=i/16$
which would have appeared in the non-minimal
Ising-like theories.   Mapping this $h=1/K$ sector back to the
parafermion theory is, however, a trivial matter in this case,
for we recognize that this $h=1/K$ sector is indeed the $A$-sector
which was our starting point in (\ref{Aintproj}).
Thus, in analogy to (\ref{Avacidstwo}), we have the unique solutions
\beq
      \,A_K^b
     ~=~  \,A_K^f
      ~=~ {{16}\over K} ~\eta^{-16/K} \, \chi_{1/K}^{(c=8/K)}
\label{rewritten}
\eeq
where the linear combinations $A_K^b$ and $A_K^f$ are
given in (\ref{Abf}).
Eq.~(\ref{rewritten}) is, of course, merely a rewriting of (\ref{Aintproj}),
and just as in the $K=2$ superstring case,
this $h=1/K$ sector indeed survives all GSO-projections
and contributes to the physical spectrum of
the fractional superstring.

It now remains only to fill out the one remaining sector
of the $c=8/K$ minimal Ising-like CFT:  this is the sector with $h=1/2$.
Proceeding as above, we find that the relevant vacuum state must
correspond to a parafermion state of highest weight
\beq
    H_1 ~=~ {2 \over{K+2}} ~+~ {1\over 2} ~-~ {1\over K}~,
\label{Htwo}
\eeq
and this can be uniquely rewritten in
terms of the parafermion highest weights as
\beq
     H_1 ~=~ (D_c-3) \,h^{K/2}_{K/2} ~+~ h^{K/2}_{K/2-2}~.
\label{Htwoagain}
\eeq
We therefore see that the corresponding
state in the parafermion theory is
\beq
    ({\rm mass})^2={1\over 2} - {1\over K}:~~~~~~~
     (\phi^{K/4}_{\pm K/4})^{D_c-3}\,
       \phi^{K/4}_{\pm(K/4-1)}~.
\label{firstexcitedstate}
\eeq
The potential projection sectors in this case
can be obtained by starting with highest weight $H'_1=H_1 +2$, for
\beq
   H'_1 ~=~ (D_c-3)\,h^K_{K/2} + h^{K-2}_{K/2-2} ~=~
            (D_c-4)\,h^K_{K/2} + h^{K-2}_{K/2} + h^K_{K/2-2}~.
\label{Hone}
\eeq
(The second expression is again appropriate only for cases with
$D_c \geq 4$.)
Thus, we have in general the two possible projection sectors
\beq
   ({\rm mass})^2= {5\over 2} -{1\over K}:~~~~~~~
            (\phi^{K/2}_{\pm K/4})^{D_c-3} \phi^{K/2-1}_{\pm (K/4-1)}~,~~~
            (\phi^{K/2}_{\pm K/4})^{D_c-4} \phi^{K/2-1}_{\pm K/4}
             \phi^{K/2}_{\pm (K/4-1)}~,
\label{Honeagain}
\eeq
and all other potential projection sectors (some of which have
$H'_1=H_1+1$) can be obtained by
substituting $\phi^j_m \leftrightarrow \phi^{K/2-j}_m$.
Constructing linear combinations of the corresponding string functions
in order to determine which of these sectors are indeed projection sectors,
we find that we can once again reproduce the characters of the $h=1/2$ sectors
of the $c=8/K$ minimal Ising-like models.
Specifically, defining
\beqn
    K=4:&~~~~~~~~~
      V_4 ~&\equiv~ {\textstyle{1\over 4}} d^2_0 (d^2_2)^3 -
      {\textstyle{3\over 2}} d^0_0 (d^4_2)^2 d^2_2-
      {\textstyle{1\over 2}} d^2_0 (d^4_2)^3  \nonumber\\
    K=8:&~~~ ~~~~~~
      V_8~&\equiv~ {\textstyle{1\over 2}} d^4_2 d^4_4 -
                             d^2_2 d^8_4 -
                             d^6_4 d^8_2 \nonumber\\
    K=16:&~~~ ~~~~~~
      V_{16}~&\equiv ~ {\textstyle{1\over 2}} d^8_6 - d^{14}_6 ~,
\label{moreids}
\eeqn
we find
\beq
       V_K ~=~ {{16}\over K}~
        \eta^{-16/K}\,\chi_{1/2}^{(c=8/K)}~.
\label{moreidstwo}
\eeq
Thus, we see that in this case, {\it all}\/ of the sectors in
(\ref{Honeagain}) serve as projection sectors.
Note that {\it unlike}\/ the $K=2$ superstring case,
this $h=1/2$ sector is once again completely GSO-projected out
of the resulting fractional-superstring spectrum.
This, of course, agrees completely with the fractional-superstring
partition functions obtained in (\ref{partfuncts}).

This, then, completes the partition-function mapping between the parafermionic
theory and the post-projection minimal Ising-like theory described in
Sect.~3.1.  Specifically, collecting our results, we have found
\beqn
          U_K ~&=&~  ~~~~~~~~\eta^{-16/K}~\chi_0^{(c=8/K)} \nonumber\\
  A_K^b=A_K^f ~&=& ~(D_c-2)~\eta^{-16/K}~\chi_{1/K}^{(c=8/K)} \nonumber\\
          V_K ~&=& ~(D_c-2)~\eta^{-16/K}~\chi_{1/2}^{(c=8/K)}~.
\label{Aresults}
\eeqn
Note the factors of $D_c-2$ which precede the characters of
the two ``excited'' sectors with $h>0$.
The suggestive appearance of (\ref{Aresults})
thus leads us to conclude that
the post-projection $A$-sector CFT for $K>2$ is
indeed the minimal $c=8/K$ Ising-like CFT tensored together with $D_c-2=16/K$
free bosons.
It is in fact straightforward to verify that there exist no other
string-function combinations which, together with internal projections,
could produce the {\it other} Ising-like characters which
would have arisen in any of the other {\it non}-minimal Ising-like
models discussed in Sect.~3.1.
It is this fact which is the ultimate justification for
our selection of the minimal Ising-like CFT's in (\ref{CFTchoice}).

\subsection{Spacetime Statistics}

The above description, however, is still incomplete, for while our
fractional-superstring {\it characters} have been related to the
 {\it characters} of the minimal Ising-like CFT, the difficult
task of finding a proper worldsheet formulation or representation
of this post-projection CFT still remains.  One indication of this
difficulty can already be seen from the above results.  Note from
(\ref{Aresults}) that we have related {\it both}\/ $A_K^b$ and $A_K^f$ to
the {\it same} character of the same minimal Ising-like CFT.  Although $A_K^b$
and $A_K^f$ are of course equal when expressed as functions of $q$ (since in
our supersymmetric theory we have $A_K=A_K^b-A_K^f=0$), we do not expect these
quantities to be equal when expressed in terms of conformal field theory
 {\it characters}, for these two distinct sectors (one spacetime bosonic and
the other spacetime fermionic) cannot both be expected to originate from
the {\it same}\/ vacuum state or set of primary fields in our underlying
post-projection CFT.  In the $K=2$ superstring case, this problem does not
arise, for although we find $A_2^f = 8 \eta^{-8} \tilde \chi_{1/2}^{(c=4)}$
[where this is the $h=1/K$ character in the second
line of (\ref{Aresults}), corresponding to the Ramond sector], we
in fact also find
$A_2^b= 8 \eta^{-8} \chi_{1/2}^{(c=4)}$
[where this is the $h=1/2$ character
in the {\it third}\/ line of (\ref{Aresults}), corresponding
to the NS sector with positive $G$-parity];  indeed, for the $K=2$
special case these two characters are equal due to the Jacobi identity
$\chi_{1/2}^{(c=4)} =\tilde \chi_{1/2}^{(c=4)}$.  Thus, for $K=2$, we
see that we should really identify $A_2^b$ with the $V_2$ sector,
and it is for this reason that
the $\chi_{1/2}^{(c=8/K)}$ sector survives the GSO-projection in the $K=2$
superstring case even though it fails to do so for the $K>2$
fractional-superstring cases.  For $K>2$, however, we see that neither the
bosonic expression $A_K^b$ nor the fermionic expression $A_K^f$ can be
identified with $V_K$, since both have the highest weight $h=1/K$.

The question then arises as to whether it is $A_K^b$ or $A_K^f$
(or some linear combination of the two)
which is to be identified with the $h=1/K$ sector of the Ising-like CFT.
This can be answered unambiguously
by employing modular transformations, however, for
although two distinct expressions such as $A_K^b$ and $A_K^f$ may be equal
as functions of $q$, they need not transform identically when expressed as
functions of their
underlying characters.\footnote{A trivial example of this fact
  appears even for the $K=2$ case:  the expressions $\chi_{1/2}^{(c=4)}$ and
  $\tilde \chi_{1/2}^{(c=4)}$ in (\ref{chars_four}) are equal as functions of
  $q$, but transform quite differently under the $S:\,\tau\to -1/\tau$ modular
  transformation.  Indeed, only their difference is truly modular-invariant.}
It is, however, a simple matter to determine the transformation properties of
the characters on
the right sides of (\ref{Aresults}), and using the definitions
for $U_K$, $A_K^b$, $A_K^f$, and $V_K$ in terms of the parafermionic
string functions, we can similarly determine the transformation
properties of these four fractional-superstring quantities which appear
on the left sides of (\ref{Aresults}).
We find, expectedly, that $U_K$ and $V_K$ transform precisely
as do their corresponding counterparts in (\ref{Aresults}),
but surprisingly it is only the linear combination
\beq
      \tilde A_K ~\equiv ~  \half \left( A_K^b \,+\,A_K^f\right)
\label{AKtilde}
\eeq
which transforms in the same way as $\chi_{1/K}^{(c=8/K)}$ for all $K>2$.
Specifically, under $T:\,\tau\to\tau+1$ and $S:\tau\to -1/\tau$ we have
\beq
    \pmatrix{  U_K \cr  \tilde A_K \cr V_K \cr }~
     {\buildrel T \over \longrightarrow}
          ~T^{(K)}_{ij} \,
     \pmatrix{  U_K \cr  \tilde A_K \cr V_K \cr } ~;~~~~~
    \pmatrix{  U_K \cr  \tilde A_K \cr V_K \cr }~
     {\buildrel S \over \longrightarrow}
          ~(-i\tau)^{-8/K}~ S^{(K)}_{ij} \,
     \pmatrix{  U_K \cr  \tilde A_K \cr V_K \cr }
\label{Amixings}
\eeq
where
\beq
     T^{(K)}_{ij}~=~ \exp\left[2\pi i(h_i - 1/K)\right] \,\delta_{ij}
\label{Tmatrix}
\eeq
and
\beq
    S^{(4)}_{ij}~=~ S^{(8)}_{ij}~=~ {1\over 4}\,\pmatrix{
     2 & 2 & 2 \cr
     4 & 0 & -4 \cr
     2 & -2 & 2 \cr};~~~~
     S^{(16)}_{ij}~=~ {1\over 2}\,\pmatrix{
     1 & \sqrt{2} & 1 \cr
     \sqrt{2} & 0 & -\sqrt{2} \cr
     1 & -\sqrt{2} & 1 \cr}~.
\label{Smatrices}
\eeq
Note, as a check, that the $S_{ij}$ matrices must in general
diagonalize the fusion rules of a given CFT \cite{Verlinde}, so that
the fusion rules
\beq
      [\phi_i]~\times~ [\phi_j] ~=~ \sum_{k}\,N_{ijk}\,[\phi_k]
\label{fusiongeneral}
\eeq
can always be obtained from the $S_{ij}$ matrices via the Verlinde formula
\beq
        N_{ijk} ~=~ \sum_n \, {{
      S_{in}\, S_{jn}\, S_{nk} }\over{ S_{1n} }}
\label{Verlindeformula}
\eeq
where $i=1$ corresponds to the identity sector with $h=0$.
Substituting (\ref{Smatrices}) into (\ref{Verlindeformula})
indeed reproduces (\ref{Isingfusions}) for each value of $K$,
with $\lbrace \bone, \phi_1,\phi_2\rbrace \leftrightarrow
\lbrace U_K, {\textstyle{{K}\over {16}}} V_K, {\textstyle{{K}\over {16}}}
\tilde A_K\rbrace$.
Thus, it is only the linear combination (\ref{AKtilde}) which
can legitimately be identified with the $h=1/K$ sector of
the $c=8/K$ minimal Ising-like CFT:
\beqn
          U_K ~&=&~  ~~~~~~\eta^{-16/K}~\chi_0^{(c=8/K)} \nonumber\\
  \tilde A_K  ~&=& ~(D_c-2)~\eta^{-16/K}~\chi_{1/K}^{(c=8/K)} \nonumber\\
          V_K ~&=& ~(D_c-2)~\eta^{-16/K}~\chi_{1/2}^{(c=8/K)}~.
\label{Aresultstwo}
\eeqn

The above result implies that the $h=1/K$ sector of our minimal
$c=8/K$ Ising-like CFT is neither purely spacetime bosonic nor
fermionic for $K>2$, but instead consists of states of both varieties!
This alone indicates the difficulty of formulating worldsheet representations
of our minimal Ising-like theory which naturally yield these appropriate
spacetime properties for $K>2$.  In the $K=2$ superstring case, for example,
we established that $A_2^f$ is to be identified with the $h=1/K$ sector,
and that $A_2^b\equiv V_2$ is to be identified with the $h=1/2$ sector.
This is consistent with our understanding that for $K=2$, the $h=1/K$ sector
is built upon the Ramond vacuum state $\sigma^8$:  this state produces the
necessary cuts for the worldsheet fermions and supercurrent, and serves
as the fundamental spinor which allows all resulting spacetime particles
to transform under the transverse eight-dimensional Lorentz group as
spacetime fermions.  Likewise, the $h=0,1/2$ sectors are
built upon vacuum states of the form $\bone^i\psi^{8-i}$, and these have
natural spacetime bosonic interpretations.  For $K>2$, however, we require
a representation for our minimal Ising-like CFT (\ref{CFTchoice}) which
yields {\it mixed}\/ spacetime statistics properties for the $h=1/K$
sector, simultaneously containing states which are vectors and spinors
under $SO(D)$.  Thus, unlike the $K=2$ superstring case, we no longer
anticipate that this minimal Ising-like CFT can be represented simply
as a tensor product of of worldsheet bosons and fermions.
This issue will be discussed further in Sect.~6, where we shall obtain
more detailed information concerning the spacetime statistics properties
of the individual states which contribute to these $h=1/K$ sectors.

For $K>2$, then, we see that the post-projection CFT of the
fractional superstring is {\it isomorphic}\/ to the minimal Ising-like
CFT in (\ref{CFTchoice}) --- meaning that these two theories share the
same central charges, highest weights, fusion rules, and characters ---
but that a new representation of this CFT
in terms of worldsheet fields will be necessary
in order to adequately describe the spacetime statistics properties
of the resulting sectors.


\section{Internal Projections in the $B$-Sector}
\setcounter{footnote}{0}

We now turn to the remaining sectors of the fractional superstring,
namely the massive $B$-sectors discussed in Sect.~2.  In particular,
we seek to subject these $B$-sectors to an analysis analogous to that
employed for the $A$-sectors in the previous section, and our corresponding
starting point will be the result (\ref{Bsplit}).
Note that we will be focusing here exclusively on the $K=4$
and $K=8$ fractional superstrings, for (\ref{partfuncts}) indicates
that the $K=2$ and $K=16$ theories contain no $B$-type sectors.

In order to interpret (\ref{Bsplit}) in terms of the {\it characters}
of some post-projection CFT, we now must interpret the {\it scaling}\/
of $\tau$, for CFT characters are generally defined only in terms
of an {\it unscaled}\/ $\tau$:
\beq
         \chi_h(\tau) ~\equiv~  {\rm Tr}_{h} \, \exp(2\pi i H\tau)~.
\label{chardeftau}
\eeq
Here $H$ is the Hamiltonian of the two-dimensional CFT,
and the trace is over the (appropriately defined) sector
of highest weight $h$.
Thus, we see that a scaling of $\tau$ can generally be reinterpreted
as a rescaling of the {\it energies}\/ in our underlying two-dimensional
worldsheet theory.  More specifically, let us recall the character
$\chi_{1/8}^{(c=1)}$ of the $c=1$ Dirac theory introduced in
(\ref{chars_one}):
\beq
     \chi_{1/8}^{(c=1)} ~\equiv~
    {{\vartheta_2}\over{2\eta}} ~=~ \eta^{-1} \,\sum_{n\in\bZ}
            \,q^{(n+1/2)^2/2}~,~~~~~~~q\equiv \exp(2\pi i\tau)~.
\label{sumrepn}
\eeq
This expression, which is the square of the Ising-model character
$\chi_{1/16}$, is usually interpreted as the character
of a single worldsheet {\it complex} fermion with periodic (Ramond)
space boundary conditions on the torus:
the $\eta$-denominator then represents the contribution
to the trace from the infinite tower of states built upon each
vacuum state in the Ramond sector,
while the summation tallies the contributions from
these (infinitely many) vacuum states.
Note that these vacuum states have worldsheet momenta $P=n+1/2$
with $n\in\IZ$, and thus they together form an internal
one-dimensional momentum lattice with lattice spacing one;
the energies of these vacuum states
are then generally given by $H=P^2/2$.  Thus, by analogy, we see that
the expression $\vartheta_2(\lambda\tau)/2\eta(\tau)$ corresponds
to a complex Ramond fermion formulated on a {\it rescaled momentum lattice}\/
with lattice spacing $\sqrt{\lambda}$;
the energy scale for the infinite oscillator tower of
states built upon each of these momentum vacuum states is then unchanged.
Note that only this {\it partial}\/
rescaling (momentum lattice rescaled but
oscillator tower untouched) yields a consistent theory, for only
this partial rescaling
is equivalent to a change in the radius of compactification
of the $c=1$ boson to which these fermion theories are equivalent.
(The connections of this $B$-sector theory to that
of a compactified boson are discussed in Sect.~5.)

Thus, (\ref{Bsplit}) indicates
that the $B$-sector internal projections appear to reduce our original
parafermion theories down to those of fermions formulated on rescaled
momentum lattices (or ``rescaled Ising models''), in the same way that
the $A$-sector internal projections appeared to reduce
these theories down to those of
fermions on {\it un}\/scaled momentum lattices (or ``unscaled Ising models'').
This result implies that the central charges of the post-projection
CFT's in the $B$-sectors are the {\it same} as those of the $A$-sectors,
for this partial rescaling (or radius change)
does not alter the underlying central charge.
Obtaining the same central charge in all sectors is of course necessary
if the fractional superstring is ultimately to be
described in terms of a single post-projection CFT.
This in turn implies that the parafermionic ``effective vacuum state''
for the $B$-sectors must be
the {\it same} as that of the $A$-sectors given in (\ref{effectivevacuum}),
yet another indication that the behavior of the
internal projections in the $B$-sectors
is consistent with that of the projections
of the $A$-sectors.

We still must determine, however, whether our $B$-sector
post-projection CFT's are isomorphic to the ``minimal'' combination
of $D_c-2$ rescaled Ising models, or perhaps to some other larger combination.
In fact, since no $B$-sectors
appear in the ordinary $K=2$ superstring case, we have no guide
as to whether the ``minimal'' assumption is correct in the one special case
(that of the ordinary superstring) whose underlying CFT is well-understood.
Hence, we shall adopt a somewhat different approach from that of
Sect.~3 in analyzing the post-projection CFT's of the $B$-sectors.
Recall that in Sect.~3,
we started by guessing that the minimal Ising-like theories
were in fact our post-projection CFT's for the $A$-sectors;
this in turn implied the existence of three sectors with highest weights
$h=0,1/K,1/2$ whose characters $U_K, \tilde A_K, V_K$
we eventually constructed out of parafermionic
string functions.  These characters were then found
to form a closed system under modular transformations,
mixing under the $S$ modular transformation
according to matrices $S_{ij}^{(K)}$ from which the
originally assumed CFT fusion rules (\ref{Isingfusions}) were verified
via (\ref{Verlindeformula}).
We shall apply this same procedure here, therefore, but in reverse.
Starting with the characters $B_K^{b,f}$ in (\ref{Bbf}) expressed as linear
combinations of string functions, we shall take modular transformations in
order to fill out the complete system of characters in our
$B$-sector post-projection CFT.  The string-function
combinations which comprise this system will thus be the
$B$-sector analogues of the three $A$-sector characters $U_K, \tilde A_K,$
and $V_K$, and will thereby provide (as before) a mapping between the sectors
of our smaller post-projection CFT and the
larger (but internally projected) original parafermion theory.  From this
set of characters we will then be able to infer the
relevant spectrum of highest weights in our post-projection CFT,
as well as its complete set of fusion rules.

It is a straightforward matter
to take the modular transformations of $B_K^{b,f}$ in (\ref{Bbf}),
for the modular transformations of the individual string functions
$c^\ell_n$ are well-known.  We find the following results.
For the $B$-sectors, we
now find that our post-projection CFT has {\it nine}
sectors for each value of $K$, since
there are nine linearly independent combinations
of string functions required
for closure under modular transformations.
In each case, however, we can simplify matters by grouping
these nine characters into three groups of three, for
we will see that each such group by itself resembles the
three-sector minimal Ising-like theory.
Thus, we choose a notation in advance which reflects this analogy:
we will denote these three minimal Ising-like theories as
copies $(a)$, $(b)$, and $(c)$, and denote the three sectors {\it within}
each copy (in analogy with those for the $A$-sector)
as $X_K, \tilde B_K,$ and $Y_K$.
Then our nine string-function combinations are as follows.
Recalling the definition $d^\ell_n\equiv c^\ell_n + c^{K-\ell}_n$,
 we find for $K=4$:
\beqn
   X_4^{(a)} ~\equiv&~ (d^0_0)^3 d^2_2  - (d^0_0)^2 d^2_0 d^4_2 ~~~~~~~~~
             ~&(h= 0)\nonumber\\
   \tilde B_4 ~\equiv~ \tilde B_4^{(a)}
      ~\equiv&~ \half(B_4^b+B_4^f)~~~~~~~~~~~~~~~~~
             ~&(h= 3/4)\nonumber\\
   Y_4^{(a)} ~\equiv&~ d^0_0 d^2_2 (d^4_2)^2 - d^2_0 (d^4_2)^3~~~~~~~~~
             ~&(h= 3/2)~ \nonumber\\
   X_4^{(b)} ~\equiv&~ (d^0_0)^2 d^2_0 d^2_2  - d^0_0 (d^2_0)^2 d^4_2 ~~~~~~~~~
             ~&(h= 1/3)\nonumber\\
   \tilde B_4^{(b)} ~\equiv&~ (d^0_0)^2 (d^2_2)^2  - (d^2_0)^2 (d^4_2)^2
   ~~~~~~~~~ ~&(h= 1/12)\nonumber\\
   Y_4^{(b)} ~\equiv&~ d^0_0 (d^2_2)^2 d^4_2  - d^2_0 d^2_2 (d^4_2)^2 ~~~~~~~~~
             ~&(h= 5/6)\nonumber\\
   X_4^{(c)} ~\equiv&~ d^0_0 (d^2_0)^2 d^2_2  - (d^2_0)^3 d^4_2 ~~~~~~~~~
             ~&(h= 2/3)\nonumber\\
   \tilde B_4^{(c)}
      ~\equiv&~ d^0_0 d^2_0 (d^2_2)^2  - (d^2_0)^2 d^2_2 d^4_2 ~~~~~~~~~
           ~&(h= 5/12)\nonumber\\
   Y_{4}^{(c)} ~\equiv&~ d^0_0 (d^2_2)^3  - d^2_0 (d^2_2)^2 d^4_2 ~~~~~~~~~
             ~&(h= 1/6)~.
\label{Bsectorsfour}
\eeqn
Similarly, for $K=8$, we have the combinations:
\beqn
  X_{8}^{(a)}~\equiv&~
    d^0_0 d^2_2 - d^2_0 d^8_2 ~~~~~~~~~~~~~~~&(h = 0) \nonumber\\
\tilde B_8 ~\equiv~  \tilde B_8^{(a)} ~\equiv&~ \half(B_8^b + B_8^f) ~
        ~~~~~~~~~~~~~~&(h = 5/8) \nonumber\\
  Y_{8}^{(a)}~\equiv&~ d^6_4 d^8_2 - d^2_2 d^8_4 ~~~~~~~~~~~~~~~&(h = 5/2)
         ~\nonumber\\
  X_{8}^{(b)}~\equiv&~ d^0_0 d^4_2 - d^4_0 d^8_2 ~~~~~
         ~~~~~~~~~~&(h = 2/5) \nonumber\\
  \tilde B_8^{(b)}~\equiv&~ d^0_0 d^4_4 - d^4_0 d^8_4 ~~~~~~~~~~~~~~~
       &(h = 1/40) \nonumber\\
  Y_8^{(b)}~\equiv&~ d^4_4 d^8_2 - d^4_2 d^8_4 ~~~~~~~~~~~~~~~
      &(h = 9/10)\nonumber\\
  X_{8}^{(c)}~\equiv&~ d^2_0 d^4_2 - d^2_2 d^4_0 ~~~~~~~~~~~~~~
       ~&(h = 8/5) \nonumber\\
  \tilde B_8^{(c)} ~\equiv&~  d^2_0 d^4_4 - d^4_0 d^6_4 ~~~~~~~
       ~~~~~~~~&(h = 9/40) \nonumber\\
  Y_8^{(c)}~\equiv&~ d^2_2 d^4_4 - d^4_2 d^6_4 ~~~~~~~~
       ~~~~~~~&(h = 1/10) ~.
\label{Bsectorseight}
\eeqn
Next to each string-function combination in (\ref{Bsectorsfour})
and (\ref{Bsectorseight}) we have indicated the highest weight
of the corresponding sector in the post-projection CFT.
These highest weights are readily determined
by expanding each corresponding string-function combination
in the form $q^\ell\sum_{n=0}^\infty a_n q^n$.
Since the quantity $\ell$ must then be the (mass)$^2$ of the
vacuum state in the corresponding sector, and since this
in turn must equal $h-c/24$ (where $c$, the central
charge of the post-projection CFT, is $24/K$ for both the
$A$- and $B$-sectors), we have $h=\ell+1/K$.

There are several things to note about the expressions in (\ref{Bsectorsfour})
and (\ref{Bsectorseight}).  We have already seen that there are
internal projections acting within the expressions $B_K^{b,f}$, and this was
the basis for central charge reduction in the $B$-sectors.  It is now
clear that indeed {\it each}\/ of the parafermion sectors listed in
these equations
contains an analogous internal projection as well;  this is of course required
for self-consistency, since these different sectors are parts of the same
post-projection CFT with reduced central charge $c=24/K$.  Thus, just as
for the $A$-sector expressions
$\tilde A_K, U_K,$ and $V_K$, each of the above sets of nine quantities
can be viewed as providing a mapping between the our residual smaller
$B$-sector post-projection CFT and the larger original parafermion CFT with
an internal projection.  In this vein, note that once again it is the $h=0$
sector which serves as the ``effective vacuum'' sector of the post-projection
CFT, with the characters corresponding to the same state
(\ref{effectivevacuum}) appearing within $X_K^{(a)}$ for each value of $K$.
Thus the $A$-sectors and $B$-sectors indeed share the same effective vacuum
state at (mass)$^2 = -1/K$, indicating (as claimed earlier) that the internal
projections in the $A$- and $B$-sectors appear to be consistent with
each other.  Note, however, that these are nevertheless {\it different}\/
internal projections, for they combine the
parafermionic projection and non-projection sectors in different
ways in order to produce the different $A$- and $B$-sector post-projection
CFT's.

The above sets of string-function combinations form closed systems under
modular transformations.  Indeed, under the $T$ modular transformation,
the nine combinations in (\ref{Bsectorsfour}) and (\ref{Bsectorseight})
respectively transform according to (\ref{Tmatrix}),
whereas under the $S$ transformation these sets separately transform
into themselves with the following mixing matrices:
\beq
     S_{ij}^{(K=4)}~=~ {1\over{6}}
    \pmatrix{ 1 & 2 & 1 & 2 & 2 & 2 & 1 & 2 & 1 \cr
    1 & 0 & -1 & 2 & 0 & -2 & 1 & 0 & -1 \cr
    1 & -2 & 1 & 2 & -2 & 2 & 1 & -2 & 1 \cr
    2 & 4 & 2 & 1 & 1 & 1 & -1 & -2 & -1 \cr
    4 & 0 & -4 & 2 & 0 & -2 & -2 & 0 & 2 \cr
    2 & -4 & 2 & 1 & -1 & 1 & -1 & 2 & -1 \cr
    4 & 8 & 4 & -4 & -4 & -4 & 1 & 2 & 1 \cr
    4 & 0 & -4 & -4 & 0 & 4 & 1 & 0 & -1 \cr
    4 & -8 & 4 & -4 & 4 & -4 & 1 & -2 & 1 \cr }
\label{Bfourmatrix}
\eeq
and
\beq
     S_{ij}^{(K=8)}~=~ {1\over{2}}
    \pmatrix{   r &   r &   r &   r &   r &   r &   r &   r &   r \cr
    2 r & 0 & -2 r & 2 r & 0 & -2 r & 2 r & 0 & -2 r \cr
      r & -  r &   r &   r & -  r &   r &   r & -  r &   r \cr
    2 r & 2 r & 2 r &   a &   a &   a & -  b & -  b & -  b \cr
    4 r & 0 & -4 r & 2 a & 0 & -2 a & -2 b & 0 & 2 b \cr
    2 r & -2 r & 2 r &   a & -  a &   a & -  b &   b & -  b \cr
    2 r & 2 r & 2 r & -  b & -  b & -  b &   a &   a &   a \cr
    4 r & 0 & -4 r & -2 b & 0 & 2 b & 2 a & 0 & -2 a \cr
    2 r & -2 r & 2 r & -  b &   b & -  b &   a & -  a &   a \cr  }
\label{Beightmatrix}
\eeq
where $r\equiv 1/\sqrt{5}$, $a\equiv (1-r)/2$, and $b\equiv(1+r)/2$.
These matrices are each written in the
basis formed by the relevant nine string-function combinations,
with the order of each basis taken to be the same as the order
in which these combinations are listed within (\ref{Bsectorsfour})
and (\ref{Bsectorseight}).
Note that these matrices square to ${\bf 1}$, as required.
Indeed, if we examine the smaller $3\times 3$ blocks
which together comprise these $9\times 9$ matrices, we see
that these submatrices are each of the same general form
(\ref{Smatrices}) which we found for the $A$-sectors in Sect.~3
(up to renormalizations of our nine quantities).
This is our first indication that our nine-sector theory can
be decomposed into three copies of a theory resembling
the three-sector minimal Ising-like theories of the $A$-sectors.

In order to rigorously define the manner in which this theory
can be viewed as three such copies, we now
proceed to determine the fusion rules of this theory.
Recall that these fusion rules can
be determined from the $S_{ij}^{(K)}$-matrices (\ref{Bfourmatrix})
and (\ref{Beightmatrix}) via the
formula (\ref{Verlindeformula}).  {\it A priori}, however,
there will be 36 linearly independent
non-trivial fusion rules for {\it each}\/ of these nine-sector systems.
Let us therefore first organize these fusion rules in a coherent
fashion.  We already suspect that we have three copies of an Ising-like
theory;  these copies are labeled $(a)$, $(b)$, and $(c)$.
Furthermore, {\it within} each copy, we expect that
$X_K$ plays the role of the identity $\bone$, while $\tilde B_K$ plays
the role of the spin field $\sigma$, and $Y_K$ plays the role of the
Majorana fermion $\psi$.
We therefore expect two kinds of fusion rules:  those which indicate
how any two copies of this Ising-like theory fuse together, and those
which indicate how the fields {\it within} each copy fuse together.
More specifically, we must determine the fusions in
$\lbrace X_K, \tilde B_K, Y_K\rbrace$ space as well as those
in $\lbrace a,b,c \rbrace$ space. Let $[\phi^{(s)}]$ generically indicate a
sector in copy ($s$) of the Ising-like theory ($s=a,b,c$).  Then our results
are as follows.  For $K=4$, the fusion rules {\it between} the copies are
as follows:
\beqn
   K=4:~~~~~~
      \lbrack \phi^{(a)}\rbrack
    \times [\phi^{(s)}] ~&=&~ [\phi^{(s)}] ~~~~~~~~(s=a,b,c)\nonumber\\
      \lbrack \phi^{(b)}\rbrack
    \times [\phi^{(b)}] ~&=&~ 4\,[\phi^{(a)}] ~+~
         [\phi^{(b)}] ~+~ 2\,[\phi^{(c)}] \nonumber\\
      \lbrack \phi^{(b)}\rbrack
    \times [\phi^{(c)}] ~&=&~
         2\,[\phi^{(b)}] ~+~ 2\,[\phi^{(c)}] \nonumber\\
      \lbrack \phi^{(c)}\rbrack
    \times [\phi^{(c)}] ~&=&~ 4\,[\phi^{(a)}] ~+~
         2\,[\phi^{(b)}] ~+~ [\phi^{(c)}] ~,
\label{Bfusionsfour}
\eeqn
whereas for $K=8$ we find instead:
\beqn
   K=8:~~~~~~
      \lbrack\phi^{(a)}\rbrack
    \times [\phi^{(s)}] ~&=&~ [\phi^{(s)}] ~~~~~~~~(s=a,b,c)\nonumber\\
      \lbrack\phi^{(b)}\rbrack \times [\phi^{(b)}] ~&=&~ 2\,[\phi^{(a)}] ~+~
          [\phi^{(c)}] \nonumber\\
      \lbrack\phi^{(b)}\rbrack
    \times [\phi^{(c)}] ~&=&~ [\phi^{(b)}] ~+~ [\phi^{(c)}] \nonumber\\
      \lbrack\phi^{(c)}\rbrack
    \times [\phi^{(c)}] ~&=&~ 2\,[\phi^{(a)}] ~+~ [\phi^{(b)}]  ~.
\label{Bfusionseight}
\eeqn
 {\it Within}\/ each Ising-like copy, however, we
indeed find that our fusion rules are the usual Ising-model
fusion rules (\ref{Isingfusions}), with
$\lbrace \bone,\varphi_1,\varphi_2\rbrace \sim
\lbrace X_K,Y_K,\tilde B_K\rbrace$.
Thus we see that we indeed have three distinct ``copies'' of an
Ising-like theory in both the $K=4$ and $K=8$ cases
(in the sense that they have the same internal fusion rules as
the Ising-like theories of Sect.~3.1),
and these cases differ only in the manner in
which these three different copies are sewn (or fused) together.
For example, putting these two types of fusion rules together yields
the fusion rule:
\beq
   \lbrack \tilde B_K^{(c)} \rbrack
   \times \lbrack \tilde B_K^{(c)} \rbrack
  ~=~ \sum_{s=a,b,c}  \, n_s \, \left( [X_K^{(s)}] + [Y_K^{(s)}] \right)~
\label{fusionexample}
\eeq
with $\lbrace n_a,n_b,n_c\rbrace =\lbrace 4,2,1\rbrace$ for $K=4$ and
$\lbrace 2,1,0\rbrace$ for $K=8$.
Note, however, that these individual copies bear no other relation to the
minimal Ising-like theories of Sect.~3.1:  they do not individually close
under fusion as do the latter minimal theories, nor do their
highest weights correspond.

The fusion rules (\ref{Bfusionsfour}) and (\ref{Bfusionseight})
are of course associative, as is guaranteed by construction since they
were obtained via (\ref{Verlindeformula}) from $S$-matrices satisfying
$S^2=\bone$.
The fact that some fusion-rule coefficients are greater than one
suggests that some of these sectors actually appear in the post-projection
CFT with multiplicities exceeding one.
This would in turn suggest that there exist conserved quantum numbers
according to which these multiple sectors might be distinguished, and
we will see in Sect.~5 that this is indeed the case.


\section{Relation to Compactified Bosons}
\setcounter{footnote}{0}

In Sects.~3 and 4 we derived some of the minimal conditions (central charges,
highest weights, and fusion rules) that must be satisfied by the
post-projection CFT's for both the fractional superstring $A$- and $B$-sectors.
In this section we will demonstrate that all of these constraints can be
reformulated naturally as the properties of the CFT's of worldsheet bosons
compactified on circles of certain radii, thereby providing a uniform language
for discussing the $A$- and $B$-sectors on the same footing.  In particular,
we will be able to provide a direct mapping between the various $B$-sectors
listed in (\ref{Bsectorsfour}) and (\ref{Bsectorseight}) and the different
winding-mode sectors of the compactified-boson theory, thereby explaining the
appearance of these additional sectors and yielding explicit $B$-sector
analogues of the identities (\ref{Aresultstwo}).  Moreover, the
compactified-boson theory will provide a useful additional quantum number
(namely the $U(1)$ charge $\alpha$) through which the infinite number
of individual states contributing to the $A$- and $B$-sectors might be
distinguished.  The results of this section will also prove vital in
Sect.~6, where we will consider the problem of spacetime statistics
at the level of individual $\alpha$-states
through the use of the so-called ``twist current''.

We remind the reader, however, that the relation between our post-projection
CFT's and the compactified-boson theories is only an isomorphism which holds
at the level of their central charges, highest weights, fusion rules,
and characters.  Indeed, as we have seen (and as will become even clearer
in Sect.~6), our post-projection CFT's cannot ultimately be represented
in terms of such free worldsheet bosonic fields, and an alternative
representation remains to be found.

In order to establish conventions and notation, we begin in Sect.~5.1 by
reviewing the compactified-boson CFT and its associated characters.
The reformulation of our above results for the $A$- and $B$-sectors
will then be given in Sect.~5.2.  Finally, Sect.~5.3 contains comments
concerning the fractional-superstring $C$-sectors, all of whose states
are removed by the internal projections.

\subsection{Compactified Bosons and their Chiral Characters}

Let us consider a free (chiral) bosonic field $\phi(z)$,
normalized in the usual fashion so that
\beq  \langle\phi(z)\phi(w)\rangle ~=~ -\ln(z-w)~,~~~~~
    T(z)~=~ -\half \,:[\partial \phi(z)]^2:~,
\eeq
and compactified on a circle of radius $R$ so that $\phi\approx \phi+2\pi R$.
It is then straightforward to demonstrate that this conformal field
theory has $c=1$, with primary fields $i\partial \phi$ of weight
$h=1$ and $\epsilon^{(\alpha)}\equiv \exp(i\alpha \phi)$ of weight
$\alpha^2/2$.  The field $\epsilon^{(0)}\equiv \bone$
thus serves as the identity,
and the fusion rules for this theory take the form
\beq     \lbrack \epsilon^{(\alpha)}\rbrack ~\times~ \lbrack
       \epsilon^{(\beta)}\rbrack ~=~
       \lbrack \epsilon^{(\alpha+\beta)}\rbrack
\label{bosonfusions}
\eeq
where we understand the sector $[\epsilon^{(0)}]$ to include $\bone$,
$i\partial \phi$, and all of its descendents.
Thus $\alpha$ appears as a conserved quantum number under fusion ---
indeed, it is the {\it charge}
of the primary field $\epsilon^{(\alpha)}$ with respect to the $U(1)$
current $i\partial\phi$.

The above results are independent of the radius of compactification.
If the radius of compactification is infinite, however,
so that $R\to\infty$, then $\alpha$ is entirely unconstrained,
whereas for finite $R$ we find that $\alpha$ is restricted to the values
\beq   \alpha_{m\ell} ~=~ {m\over{2R}} ~+~ \ell R~,~~~~~m,\ell\in\IZ~.
\eeq
Here $m$ and $\ell$ respectively represent the boson momentum-
and winding-mode quantum numbers of the corresponding state
$|m,\ell\rangle \equiv \epsilon^{(\alpha_{m\ell})}(0)|0\rangle$,
and it is clear that this selected set of permitted values
for $\alpha_{m\ell}$ yields
a consistent subalgebra of our fusion rules for any $R$.
We may rewrite this set of values as follows.
Restricting ourselves to those radii for which this conformal
field theory is rational --- namely, $R^2\in {\bf Q}$ ---
we can without loss of generality write $R=\sqrt{a/(2b)}$
where $a,b$ are positive, relatively prime integers, and we
define $N\equiv 2ab$.  We then see that
\beq   \alpha_{m\ell} ~=~ {{mb+\ell a}\over{\sqrt{N}}} ~,
\eeq
and since $r\equiv mb+ \ell a\in \IZ$, we find that allowed
values for $\alpha_{m\ell}$
for arbitrary radius $R$ are simply $r/\sqrt{N}$, $r\in \IZ$.
The conformal dimensions for these primary fields $\epsilon^{(\alpha)}$
are therefore given by
\beq   h_{m\ell} ~=~  {{(\alpha_{m\ell})^2}\over 2} ~=~
         {{r^2}\over {2N}}~=~
         {{(mb+\ell a)^2}\over {2N}}~.
\eeq

It is now a simple matter to construct the set of chiral
characters corresponding to this compactified-boson theory.
The contribution arising from a single sector $\alpha_{m\ell}$
is as usual\footnote{
    This expression assumes the absence of null states in the
    corresponding Verma module, which in turn requires $h_{m\ell}\not\in\IZ/4$.
    At the values of compactification radius $R$ that we will be
    dealing with, however, this is indeed the case for all sectors
    with non-zero highest weights.  The identity sector $[\bone]$,
    by contrast, always contains a null state at level one
    (corresponding to the primary field $i\partial \phi$ with $h=1$).
    In this case (\ref{singlesector}) represents the {\it sum}
    of the contributions from both $[\bone]$ and $[i\partial \phi]$.}
\beq    Z_\alpha ~\equiv~ Z_{(m,\ell)}~=~ \eta^{-1} \, q^{h_{m\ell}}
          ~=~ \eta^{-1} \, q^{N(r/N)^2/2}~;
\label{singlesector}
\eeq
here the factor of $\eta^{-1}$
reflects the contribution from the infinite tower of states
built upon the vacuum $|m,\ell\rangle$.
Thus, writing $r=Nn+k$ where $n\in \IZ$ and $0\leq k <N$, we see
that we can construct $N$ distinct compactified-boson characters
\beq  \chi_{N,k} ~\equiv~ \eta^{-1} \, \sum_{n\in \bZ} q^{N(n+k/N)^2/2}~,
 ~~~~~~ 0 \leq k < N
\label{Chidef}
\eeq
by adding together those contributions $Z_\alpha$ from
sectors $\alpha=(m,\ell)$ satisfying
\beq     \sqrt{N}\,\alpha  ~=~ mb+\ell a~=~ k ~~~~~~({\rm mod}~N)~.
\eeq
Explicitly, we have
\beq    \chi_{N,k} ~=~ \sum_{\sqrt{N}\alpha = k\,({\rm mod}\,N)} Z_\alpha~=~
   \sum_{mb+\ell a = k \,({\rm mod}\,N)} Z_{(m,\ell)}~.
\label{alphaconstraint}
\eeq
It is for this reason that the number of
``characters'' $\chi_{N,k}$ in our compactified-boson theory is finite,
even though the number of primary fields is infinite.\footnote{
   Recall, in this context, the footnote in Sect.~3.1.  The
   compactified-boson theory thus provides us with a means of
   explicitly distinguishing the infinitely many primary fields
   in these $c\geq 1$ theories via the $\alpha$ quantum number.}

These compactified-boson characters $\chi_{N,k}$ satisfy a number
of identities.
First,
they transform covariantly under the modular group, $\chi_{N,k}(M\tau)
=\sum_{k'} M_{kk'} \chi_{N,k'}(\tau)$ where
\beq
    S_{kk'} ~=~ {1\over\sqrt{N}} \exp\left( 2\pi i \,{kk' \over N}\right)~,~~~~
   ~~
    T_{kk'} ~=~ \exp\left\lbrack 2\pi i
     \left({k^2\over 2N} - {1\over 24}\right) \right\rbrack~\delta_{kk'}~.
\label{modmatrices}
\eeq
Furthermore, we observe that $\chi_{N,k}=\chi_{N,-k}=\chi_{N,N+k}$,
and thus the set of truly independent characters is simply $\lbrace
\chi_{N,k}, ~0\leq k\leq N/2\rbrace$.
Using (\ref{Chidef}), we can also define
the corresponding (classical Jacobi-Riemann) $\Theta$-functions via
\beqn
     \Theta_{N,k} ~&&\equiv ~ \eta \, \chi_{N,k}~ =~
            \sum_{n=-\infty}^\infty \, q^{N(n+k/N)^2/2}~
      \nonumber\\
        &&=~ q^{k^2/2N}\,\prod_{n=1}^\infty \,
        (1+q^{Nn-N/2+k})\, (1+q^{Nn-N/2-k})\, (1-q^{Nn})~.
\eeqn
These $\Theta$-functions
are then related to the Jacobi $\vartheta$-functions
as follows:
\beq   \Theta_{N,k} ~=~ \cases{
      \vartheta_2(N\tau) ~=~ \left[
      \vartheta_3(N\tau/4) - \vartheta_4(N\tau/4)\right]/2 & if $k=N/2$\cr
      \vartheta_3(N\tau) ~=~ \left[
      \vartheta_3(N\tau/4) + \vartheta_4(N\tau/4)\right]/2 & if $k=0$ \cr
      \phantom{\vartheta_3(N\tau) ~=~}
    \left[\vartheta_2(N\tau/4)\right]/2 & if $k=N/4$ and $N\in 4{\bf Z}$~.\cr}
\label{Thetatheta}
\eeq
In particular, note the appearance of $\vartheta$-functions
with scaled arguments.

As required,
these chiral characters $\chi_{N,k}$ can be combined
with their complex conjugates in order to produce the full modular-invariant
partition function corresponding to a compactified boson at radius $R$:
\beq   Z(R) ~=~ {1\over{\eta\overline{\eta}}} \, \sum_{m,\ell \in\bZ}
       \qbar^{(m/2R-\ell R)^2/2} \, q^{(m/2R+\ell R)^2/2} ~.
\label{ZofR}
\eeq
Specifically,
defining $a$, $b$, and $N$ as above,
we then find a pair of integers $a',b'$ so that $\det {{a'~b'}\choose{a~b}}=1$,
and define $s\equiv a'b+b'a$ (mod $N$), with
$0\leq s < N$.
Note that while $a',b'$ are not uniquely determined by this
procedure, $s$ is uniquely defined.
Then it is a simple matter to see that (\ref{ZofR})
is given by
\beq   Z(R)~=~ \sum_{k=0}^{N-1} ~ \chi_{N,k}(q)\,\overline{\chi_{N,sk}(q)}~,
\label{ZofRtwo}
\eeq
which demonstrates that $\chi_{N,k}$ are indeed the proper
chiral characters of the compactified-boson theory.

We conclude this brief review with an important comment concerning
the identification of the physical radius $R$ on the
basis of a set of chiral characters $\lbrace \chi_{N,k}$, $0\leq k < N\rbrace$.
It turns out that there are three distinct types of ambiguities
which may arise, only some of which shall concern us.
First, of course, there is
the duality transformation $R\to 1/(2R)$ which is an exact symmetry
of the compactified-boson theory:
physically this interchanges momentum-modes and winding-modes, and
mathematically we see that
$N$ and $\alpha$ are invariant under this transformation, while
$(a,m)\leftrightarrow (b,\ell)$.
Thus, a given value of $N$ corresponds to either $R$ or $1/(2R)$.
This type of ambiguity shall not concern us, however, since both radii
correspond to the same theory with the same set of physical states.

The second ambiguity is more subtle.
Note that multiplying the radius $R$ of the compactified
boson theory by an integer $n$ (or equivalently dividing its dual
$\tilde R$ by $n$) produces a theory related to the original theory
by the introduction of additional momentum-mode sectors and the simultaneous
removal of corresponding winding-mode sectors (or vice versa).
For integer $n$, however, some of these additional winding-mode sectors
effectively replace the momentum-mode states which had been removed,
with the result that the entire original theory is ``embedded''
in the new (larger) theory and in fact comprises a self-consistent
subset of this larger theory.
In terms of the above characters, we find that
for $R\to R'\equiv nR$, we have $N\to N'\equiv n^2 N$, with
a particular sector $k$ of highest weight $h=k^2/(2N)$
in the original theory now described in the new theory
as that with $k'\equiv nk$.
Linear combinations of the characters $\chi_{N',k}$ in the $N'$ system
can then reproduce each of the characters $\chi_{N,k}$ of the original
system.  As an explicit example which will be relevant later, let us consider
the cases where $R=n$ with $n\in \IZ$; these theories then have $N= 4n^2$.
Those theories corresponding to smaller values of $n$
can therefore be equivalently described as
closed subsets of those corresponding to larger values,
and the relations between their associated characters are:
\beqn
     \chi_{4,0} ~&&=~ \chi_{16,0} + \chi_{16,8} ~=~
        \chi_{36,0} + \chi_{36,12} + \chi_{36,24}~=~...
      \nonumber\\
     \chi_{4,1} ~&&=~ \chi_{16,2} + \chi_{16,6} ~=~
        \chi_{36,3} + \chi_{36,9} + \chi_{36,15}~=~...
      \nonumber\\
     \chi_{4,2} ~&&=~ \chi_{16,4} \phantom{+ \chi_{16,6}} ~=~
        \chi_{36,6} + \chi_{36,18} + \chi_{36,30} ~=~....
\label{intrescaling}
\eeqn
We will require that our results be invariant under
this type of integer radius rescaling.

Finally, there exists a third type of ambiguity in identifying
the radius $R$ on the basis of the chiral characters alone:
as we have seen in (\ref{ZofRtwo}), the radius is also determined
in part by the manner in which these chiral characters are joined in
the full left/right partition function.
In the absence of a full partition function, therefore,
we shall generally assume that a given set of characters
$\lbrace\chi_{N,k}\rbrace$ corresponds to that radius
for which the corresponding partition function is diagonal.
This then amounts to the choice $R=\sqrt{N}/2$, which will be used
for discussion purposes.  None of our results, however, will depend
on this particular choice.

\subsection{The Post-Projection CFT's as Compactified Bosons}

We now turn to the fundamental issue:  that of relating our $A$- and
$B$-sector post-projection CFT's to those of compactified bosons.

As expected from two-dimensional boson/fermion equivalence,
there exists a natural relationship between our Ising-like theories
formulated on rescaled momentum lattices, and free bosons compactified on
circles of arbitrary radius.  Explicitly, from (\ref{Thetatheta}), we have
in general
\beq   \vartheta_2(\lambda\tau)~=~2\,\Theta_{4\lambda,\lambda}(\tau) ~,
\label{Ramondcirclesector}
\eeq
and thus it is evident that $\vartheta_2(\lambda \tau)$ is indeed the character
of a certain winding-mode sector of a boson compactified on
a circle of radius $R=\sqrt{\lambda}$.
Further integer scalings of $R$ are of course also possible.
This result indicates that it is possible to
relate the fractional-superstring sectors found in previous sections
to their compactified-boson counterparts for $K=4$ and $K=8$.
(For $K=16$ our light-cone post-projection CFT is
only that of the $A$-sector:  a single coordinate boson tensored
with the $c=1/2$ Ising model.  Thus of course no bosonized
description is possible.)

Let us concentrate first on the $A$-sectors.  For the $K=8$ case,
we have found that the post-projection CFT in the $A$-sector
contains three sectors
denoted $\tilde A_8$, $U_8$, and $V_8$, and
as discussed in Sect.~3, this theory is indeed that of the
Dirac fermion.  The Dirac fermion theory, however, is
equivalent to that of a boson compactified on a circle of radius $R=1$.
Thus we can easily relate their corresponding characters:
\beqn
     U_8 ~&&=~ \eta^{-2}\, \chi_{4,0}
       \nonumber\\
     \tilde A_8 ~&&=~ 2\,\eta^{-2}\, \chi_{4,1}
       \nonumber\\
     V_8 ~&&=~ \eta^{-2}\, \chi_{4,2} ~.
\label{UAVeight}
\eeqn
Similarly, for the $K=4$ case, the post-projection CFT again
contains three sectors, and as discussed this
theory is the ``minimal'' Ising-like diagonal subset of the (Dirac)$^2$ theory.
Using the characters given in Sect.~3, we find that this too can be
easily re-expressed in terms of {\it products}\/ of the compactified-boson
characters:
\beqn
     U_4 ~&&=~ \eta^{-4}\, \left[(\chi_{4,0})^2 + (\chi_{4,2})^2\right]
       \nonumber\\
     \tilde A_4 ~&&=~ 4\,\eta^{-4}\, (\chi_{4,1})^2
       \nonumber\\
     V_4 ~&&=~ 2\,\eta^{-4}\, \chi_{4,0}\,\chi_{4,2} ~.
\label{UAVfour}
\eeqn
The pattern of these products of characters thus indicates
the ``diagonal'' nature of this subset theory relative to
the full two-boson tensor-product theory.  An important point to which we will
return in Sect.~7, however, is the fact that these fractional-superstring
sectors are related to products of the {\it same} $N=4$
compactified-boson theory regardless of the value of $K$.

We now turn, however, to the corresponding $B$-sectors --- indeed,
it is for interpreting the $B$-sectors that the compactified-boson language
is especially appropriate.
In Sect.~4, it was demonstrated that for each value of $K$
these $B$-sector CFT's each contain nine sectors:
these could be organized into three groups of three each,
with corresponding characters denoted  $\tilde B_K^{(s)}$,
$X_K^{(s)}$, $Y_K^{(s)}$ where the index $s=a,b,c$ denotes
the particular group.
The explicit definitions of these characters in terms of
the parafermionic string functions were given in (\ref{Bsectorsfour})
and (\ref{Bsectorseight}).  For the $K=8$ case,
we find that these can now be simply identified as the characters
of the $N=20$ compactified-boson system:
\beqn
            X_8^{(a)} ~&&=~ \eta^{-2} \,\chi_{20,0}
       \nonumber\\
    \half\, \tilde B_8^{(a)}~&&=~ \eta^{-2} \,\chi_{20,5}
       \nonumber\\
            Y_8^{(a)} ~&&=~ \eta^{-2} \,\chi_{20,10}
       \nonumber\\
    \half\, X_8^{(b)} ~&&=~ \eta^{-2} \,\chi_{20,4}
       \nonumber\\
 \half\, \tilde B_8^{(b)} ~&&=~
          \eta^{-2} \,\left(\chi_{20,1}+\chi_{20,9}\right)
       \nonumber\\
    \half\, Y_{8}^{(b)} ~&&=~ \eta^{-2} \,\chi_{20,6}
       \nonumber\\
    \half\, X_8^{(c)} ~&&=~ \eta^{-2} \,\chi_{20,8}
       \nonumber\\
 \half\, \tilde B_8^{(c)} ~&&=~
          \eta^{-2} \,\left(\chi_{20,3}+\chi_{20,7}\right)
       \nonumber\\
    \half\, Y_8^{(c)} ~&&=~ \eta^{-2} \,\chi_{20,2}  ~.
\label{eightidentities}
\eeqn
These results thus constitute new identities relating the characters
of $\IZ_K$ parafermions and the characters of compactified bosons;
they are the $B$-sector analogues of the
identities (\ref{Aresultstwo}) found for the $A$-sectors.

Similarly, for the $K=4$ case, it was shown in Sect.~4 that
the post-projection CFT in the $B$-sector is equivalent
to a certain combination of {\it two} scaled-lattice fermions
with $\lambda=3$ (tensored together with, of course,
four {\it un}\/compactified coordinate bosons).
Thus, the $K=4$ characters can now also be simply identified
as products of the characters of the $N=12$ compactified-boson system,
and indeed we find the results
\beqn
     \half \,X_4^{(a)}~&&=~ \eta^{-4}\,[(\chi_{12,0})^2
          + ( \chi_{12,6})^2]
       \nonumber\\
     \quart\,\tilde B_4^{(a)}~&&=~ \eta^{-4}\,( \chi_{12,3} )^2
       \nonumber\\
     \quart\,Y_4^{(a)}~&&=~ \eta^{-4}\,\phantom{(}\chi_{12,0}\chi_{12,6}
       \nonumber\\
     \quart \,X_4^{(b)}~&&=~ \eta^{-4}\,[(\chi_{12,2})^2
          + ( \chi_{12,4})^2]
       \nonumber\\
     \quart \,\tilde B_4^{(b)}~&&=~ \eta^{-4}\,
          \left(\chi_{12,1}+\chi_{12,5}\right)^2~
       \nonumber\\
     \eig\,Y_4^{(b)}~&&=~ \eta^{-4}\,\phantom{(}\chi_{12,2}\chi_{12,4}
       \nonumber\\
     \eig\,X_4^{(c)}~&&=~ \eta^{-4}\,\left( \chi_{12,0}\chi_{12,4}
            + \chi_{12,2}\chi_{12,6} \right)
       \nonumber\\
     \eig\,\tilde B_4^{(c)}~&&=~
\eta^{-4}\,\left(\chi_{12,1}+\chi_{12,5}\right)
                 \chi_{12,3}
       \nonumber\\
     \eig\,Y_4^{(c)}~&&=~ \eta^{-4}\,\left( \chi_{12,0}\chi_{12,2} +
             \chi_{12,4}\chi_{12,6}\right) ~.
\label{fouridentities}
\eeqn
Once again these results represent a set of new character identities.

The above results provide especially natural interpretations
for the $A$- and $B$-sector fusion rules (\ref{Isingfusions}),
(\ref{Bfusionsfour}), and (\ref{Bfusionseight}) found in Sects.~3 and 4.
Recall that a given character $\chi_{N,k}$ corresponds to that
sector of the compactified-boson Fock space consisting of
vacuum states with $U(1)$ charges $\alpha$ satisfying (\ref{alphaconstraint}).
Thus, we see that the Ising-model fusion rules
in the $K=8$ $A$-sector arise naturally
from (\ref{UAVeight})
as the result of $U(1)$ charge conservation (\ref{bosonfusions})
in the compactified-boson theory, provided that we substitute
\beq
  \chi_{N,k} ~\longrightarrow~\half\left(\chi_{N,k}+\chi_{N,-k}\right)
\label{subone}
\eeq
in each identity above.  While such substitutions do not
affect the validity of our identities as functions of $q$ (since
$\chi_{N,k}=\chi_{N,-k}$),
we now see that they are necessary in order to accurately
reflect the physical {\it states} which contribute to each sector.
[This substitution is analogous to that in Sect.~3.3, where it was
found that $A_K^{b,f}$ in (\ref{Aresults}) should be replaced by
$\tilde A_K$.]
Similar results hold for the $K=8$ $B$-sector.
Additionally, for the $K=4$ case,
we see that $U(1)$ charge conservation
naturally yields the $K=4$ fusion rules
provided that we also {\it symmetrize} our above two-boson expressions:
\beq
 \chi_{N,k_1}\chi_{N,k_2}~\longrightarrow~
    \half \left(\chi^{(1)}_{N,k_1} \chi^{(2)}_{N,k_2}  +
         \chi^{(1)}_{N,k_2} \chi^{(2)}_{N,k_1}\right)
\label{subtwo}
\eeq
where the superscripts indicate the relevant boson system.
Thus, all of the fusion rules determined in Sects.~3 and 4
can be understood as the result of $U(1)$ charge conservation
in the isomorphic compactified-boson theory, and the multiplicities
found in (\ref{Bfusionsfour}) and (\ref{Bfusionseight}) are now
easily accounted for.  Indeed, for the $K=8$ $B$-sector, we see that
Ising copy $(a)$ consists of those compactified-boson sectors with
$\alpha=0$ (mod 5), whereas copy $(b)$ contains sectors with
$\alpha=\pm 1$ (mod 5), and copy $(c)$ contains sectors with
$\alpha=\pm 2$ (mod 5).
The $K=4$ $B$-sector is then the similar ``minimal''
combination of two $N=12$ theories which preserves
this same fusion-rule structure.
The major difference between the $A$- and $B$-sectors
is the fact that the $B$-sector radius depends on the
Ka\v{c}-Moody level $K$ of the theory, whereas the $A$-sector
radius is fixed:
\beq
     R_A~=~1~,~~~~~ R_B~=~\sqrt{\half(K+2)}~.
\label{radii}
\eeq

\medskip
\subsection{Internal Projections and the $C$-Sectors}

Until now we have had little to say about the $C$-sectors
of the fractional superstring.
Before concluding this section, therefore, we shall briefly
discuss the con\-formal-field-theoretic significance
of the fact that the $C$-sectors contain no physical states.

Although we have seen in (\ref{ABCvanish})
that $A_K$, $B_K$, and $C_K$ all vanish, this was interpreted as the
result of spacetime supersymmetry:  for the $A$- and $B$-sectors,
this does not imply the absence of bosonic and fermionic states,
but merely the equality of their numbers.  It is for this physical reason that
one cannot simply ``ignore'' the $B$-sectors, or use the result $B=0$
to claim that no such terms need appear in the partition function.
According to (\ref{Cbf}), however, the $C$-sectors contain no
states of either spacetime statistics, for {\it all}\/ of the
physical states in this sector are apparently removed by the internal
projections.  One wonders, therefore, whether it is consistent to set
each $C_K$ to zero in the partition functions (\ref{partfuncts}), and
to ignore the $C$-sectors altogether.

This question can also be phrased more mathematically.  The $B$- and
$C$-sectors were originally discovered via the construction of the
modular-invariant partition functions (\ref{partfuncts}), for it was found
via the modular transformation properties of the parafermionic string functions
that the terms $B_K$ and $C_K$ are necessary in order for each
$\calZ_K(\tau)$ to be invariant under $S:\,\tau\to -1/\tau$  and
$T:\,\tau\to \tau+1$.  However, since $A_K$, $B_K$, and $C_K$ are each
vanishing modular functions of $\tau$, each of these terms is clearly modular
invariant by itself.  In what sense, then, does modular invariance
require the presence of the $B$- or $C$-sectors when forming $\calZ_K$?

On a mathematical level, the answer to this question concerns not the modular
group {\it per se}, but rather the {\it representations}\/ of the modular
group.
In general, starting from a given set of characters $\chi_i$, one
determines the matrices $S_{ij}$ and $T_{ij}$ which describe their
mixings under the $S$ and $T$ modular transformations.  These matrices
clearly form a representation of the modular group, and satisfy
the modular-group defining relations $S^2=(ST)^3=\bone$.  One
then takes products of these characters in order to construct
a partition function which is not merely modular invariant as a function
of $\tau$, but which is also consistent with
the original modular group {\it representation}
specified by $S_{ij}$ and $T_{ij}$.
Physically, this is tantamount to saying that our partition functions,
though modular invariant, must also be consistent with an underlying
CFT interpretation.

This then explains the difference between the $A$ or $B$-sectors
and the $C$-sectors.  With respect to the {\it pre-projection}\/ CFT,
namely the parafermion CFT with its associated string-function characters,
only the full partition functions $\calZ_K$ are self-consistent,
and one cannot drop any individual term.
The internal projections, however, not only remove the $C$-sector
states:  they also change the underlying worldsheet CFT's.
With respect to these new {\it post-projection}\/ CFT's, therefore,
it is self-consistent to drop the $C_K$ expressions entirely.  Indeed,
one of the important properties of these smaller post-projection
CFT's is precisely that they furnish us with alternative
representation matrices $S_{ij}$ and $T_{ij}$ in terms of which
the $C$-sectors are completely decoupled under modular
transformations.  These are of course the matrices
(\ref{modmatrices}) associated with the various chiral compactified-boson
characters.

Thus our compactified-boson picture is fully consistent with
the absence of states in the $C$-sectors, furnishing
us with representation matrices $S_{ij}$ and $T_{ij}$ with respect to
which the absence of $C$-sector terms in the partition function
causes no inconsistency.
One unfortunate consequence of this picture, however, is the fact that
the $A$- and $B$-sector theories are apparently decoupled,
corresponding to different compactified-boson theories
with different radii of compactification.
Several approaches towards dealing with this problem will be
outlined in Sect.~7.


\section{Spacetime Statistics and the Twist Current}
\setcounter{footnote}{0}

We now turn to the so-called ``twist currents'' introduced independently
in \cite{CR} and \cite{AD}.  As we shall see, these twist currents turn out
to be of far-reaching importance in both the pre-projection and
post-projection CFT's.

These twist currents appear in a number of different ways.  First, as noted
in \cite{CR,AD}, they effect a reorganization of the fractional-superstring
Fock space during the internal projections, playing a role reminiscent of
that played by the ``screening operators'' in the Coulomb gas construction of
the Virasoro minimal models from a free boson theory.  Furthermore, as
demonstrated in \cite{CR}, they serve as the basis of an alternative derivation
of the fractional-superstring partition functions (\ref{partfuncts}).  Here,
however, we shall be using these currents as tools for determining some of the
spacetime statistics properties of the various sectors of our post-projection
CFT's.

In particular, we will begin by showing that the actions of the twist currents
are symmetries of the entire post-projection CFT's which we have constructed
in Sects.~3 and 4.  This suggests that although these currents have thus far
been constructed only in terms of the primary fields ({\it i.e.}, the
parafermion fields) of the {\it pre}\/-projection CFT, they might be described
directly in terms of the the primary fields of the post-projection
compactified-boson CFT as well.  We shall show that this is indeed the case,
and using the bosonized formalism of Sect.~5 we shall find that these twist
currents are in fact isomorphic to certain primary fields in the
compactified-boson theory.  This will in turn allow us to cast our
fractional-superstring partition functions in lattice-like language
({\it i.e.}, as a sum over the sites of an internal shifted momentum lattice)
in such a way that spacetime supersymmetry is naturally incorporated,
 {\it no}\/ internal projection remains, and the spacetime statistics of the
sectors are automatically taken into account in the usual way by a lattice
shift vector (or ``statistics vector'').  These results, while indicating a
certain self-consistency for the fractional-superstring spacetime statistics
assignments, will also dramatically illustrate some of the technical
difficulties involved in constructing representations of these
post-projection CFT's.
These results can then hopefully serve as a guide
in any future construction.

\subsection{The Twist Current as Parafermion Primary Field}

We begin by briefly reviewing the twist current and its properties.
In determining how our original parafermionic worldsheet CFT is
projected down to smaller effective CFT's in both the $A$- and $B$-sectors,
we saw that it was necessary to build a mapping between the respective
highest-weight sectors of these two theories.  In particular,
this entailed determining those $\IZ_K$ parafermion sectors which
contributed directly ({\it i.e.}, additively)
to the fractional-superstring Fock space,
and those which served as corresponding ``projection sectors'' whose states
were {\it subtracted}\/ from  (rather than added to) this Fock space.
It turns out that there exists a simple rule for determining
which sectors serve as projection sectors, given those that serve
as non-projection sectors.  Looking at the definitions of $A_K^{b,f}$
and $B_K^{b,f}$ given in (\ref{Abf}) and (\ref{Bbf}) for $K>2$, we see
that replacing $d^\ell_n\to d^\ell_{n+K/2}$ in each string-function
combination $d^\ell_n$ maps terms occurring with positive coefficients
to those with negative coefficients and {\it vice versa}.
(Recall that $d^\ell_{n+K}=d^\ell_n$.)
Indeed, this pattern is even slightly more involved,
for this substitution transforms the {\it non-projection}\/
sectors of the spacetime {\it bosonic}\/ expressions $A_K^{b}$ and $B_K^b$
into the {\it projection}\/ sectors of the spacetime {\it fermionic}
expressions $A_K^{f}$ and $B_K^f$, and {\it vice versa}.
Thus this substitution not only exchanges projection sectors with
non-projection sectors, but induces a spacetime-statistics flip as well.

What operator in the parafermionic tensor-product CFT could have this effect
under fusion?  From the fusion rules (\ref{parafusions}), it is
a simple matter to see that this operation $d^\ell_n\to d^\ell_{n+K/2}$
is the result of a fusion with the parafermion field $\phi^0_{K/4}$.
Thus we introduce the general ``twist current'' \cite{CR,AD}
\beq
   \Psi_K ~\equiv~  {\mathop\otimes _{\mu=1}^{D_c-2}} ~(\phi^{0}_{K/4})^\mu~
\label{twistcurrentdef}
\eeq
which has this effect under fusion in both the $A$- and $B$-sectors.
This operator is called a twist current for several reasons,
among them the fact that the parafermionic fields $\phi^0_m$ are in general
often referred to as the ``currents'' in the $\IZ_K$
parafermion theory, and the fact that currents of this form
(once tensored together with a suitable anti-holomorphic counterpart)
can be used in general to ``twist'' a modular-invariant partition function
in order to generate new modular-invariant combinations.
Note that this current is {\it not}\/ a simple generalization
of the spin-field $\sigma^8=(\phi^{1/2}_{\pm 1/2})^8$
which would seem to play this role in the $K=2$ case;
indeed, for $K=2$, this twist current
in (\ref{twistcurrentdef}) does not even exist (since there are no
fields $\phi^0_{1/2}$ with $j-m\not\in \IZ$).
Rather, we will see in Sect.~6.2
that the current (\ref{twistcurrentdef}) appears
as a generalization of the appropriate $K=2$ twist current
only when it is expressed
in terms of the primary fields of the {\it post-projection}\/ CFT.

Let us now investigate the general properties of this parafermionic
twist current $\Psi_K$, starting from the fusion rules (\ref{parafusions}).
Since each factor of $\phi^0_{K/4}$ in (\ref{twistcurrentdef})
corresponds to a different spacetime dimension and hence functions
independently under the fusion rules, we shall concentrate on only
a given single component $\phi^0_{K/4}=\phi^{K/2}_{-K/4}$.
It is simple to see under the fusion rules that
$(\phi^0_{K/4})^2 = \phi^{K/2}_0$, and thus in general
$(\phi^0_{K/4})^2$ is {\it not}\/ the identity.  Rather, operating
on a given parafermion field $\phi^j_m$, we find that
\beq
   (\phi^0_{K/4})^2 :~~~~~~~~  \phi^j_m ~\longleftrightarrow~ \phi^{K/2-j}_m~.
\label{currentsquared}
\eeq
Thus, under $(\phi^0_{K/4})^2$, only the combination of sectors
$[\phi^j_m] + [\phi^{K/2-j}_m]$ is invariant.  Remarkably, however,
all of the parafermion sectors which appear in our fractional superstrings
do so in precisely these combinations, and indeed all of the partition
functions for $K>2$ can be expressed solely in terms of the $d^\ell_n\equiv
c^\ell_n+c^{K-\ell}_n$ combinations.
This is ultimately a consequence of the fact that the pre-projection
fractional-superstring worldsheet theory is constructed using only
that subset of the $\IZ_K$ parafermion theory involving the
integer-spin fields $\phi^j_m$.
Thus, for all sectors appearing in our fractional superstrings,
we find that $(\phi^0_{K/4})^2$ functions as the identity, and indeed
we can associate
\beq
         (\Psi_K)^2 ~\Longleftrightarrow~ \bone~.
\label{currentsquaredidentity}
\eeq
This is clear for the $A_K$ and $B_K$ sectors
in (\ref{Abf}) and (\ref{Bbf}),
where we explicitly have
\beqn
  \Psi_K :~~~~~~&&  A_K^{b,f} ~\longleftrightarrow~ -A_K^{f,b}~ \nonumber\\
  &&  B_K^{b,f} ~\longleftrightarrow~ -B_K^{f,b}~,
\label{PsionAB}
\eeqn
which implies
\beqn
  \Psi_K :~~~~~~&&  \tilde A_K ~\longrightarrow~ -\tilde A_K~ \nonumber\\
                && \tilde B_K ~\longrightarrow~ -\tilde B_K~.
\label{PsionABtilde}
\eeqn
The relative minus sign introduced by each application of $\Psi_K$
is simply the reflection of the exchange
of projection and non-projection sectors.

In Sects.~3 and 4 we found that $\tilde A_K$ and $\tilde B_K$
each correspond to merely one sector of their respective fractional-superstring
post-projection CFT's.
It is thus natural to ask whether the action of $\Psi_K$
appears as a general symmetry of
the {\it entire} post-projection theories.
We have already determined that the two remaining
sectors of our $A$-sector post-projection theory
are $U_K$ and $V_K$, however, and their character definitions
are given in (\ref{Avacids}) and (\ref{moreids}).
Under the twist current, then, we indeed find
\beq
       \Psi_K :~~~~~~ U_K ~\longleftrightarrow~ - V_K~,
\label{PsiUV}
\eeq
and thus in analogous fashion we interpret $U_K$ and $V_K$ as corresponding
to sectors of opposite spacetime statistics.
(We stress that unlike $A_K^b$ and $A_K^f$,
these additional sectors $U_K$ and $V_K$ are not spacetime superpartners
of each other;  in particular, they have different highest weights and
contain states at unequal mass levels.)
We therefore see that the action of $\Psi_K$ is in fact a symmetry
of the entire $A$-sector post-projection CFT:
\beqn
    \Psi_K :~~~~~~
    &&~ \tilde A_K ~\longleftrightarrow~ - \tilde A_K \nonumber\\
    &&~ U_K ~\longleftrightarrow~ - V_K~.
\label{twisteffectA}
\eeqn

We find similar results for the $B$-sector post-projection CFT,
which was shown in Sect.~4 to contain a total of nine sectors.
These nine sectors were denoted $U_K^{(s)}, \tilde B_K^{(s)}$, and $V_K^{(s)}$
for $s=a,b,c$, and their characters were given in terms of string
functions in (\ref{Bsectorsfour}) and (\ref{Bsectorseight}).
Under the twist current, we likewise find
\beqn
   \Psi_K:
       &&~~~~~~~ \tilde B_K^{(s)} ~\longleftrightarrow~ -\tilde B_K^{(s)}
         \nonumber\\
        &&~~~~~~~  X_K^{(s)} ~\longleftrightarrow~ - Y_K^{(s)} ~~~~~~~~
        (s=a,b,c)~,
\label{twisteffectB}
\eeqn
and thus we see that the effect of the twist current
in the $B$-sectors is completely analogous to that for the $A$-sectors,
with each $B$-sector Ising-like copy transforming under $\Psi_K$
precisely as do the minimal Ising-like $A$-sector theories,
and with projection and non-projection sectors interchanged by $\Psi_K$.

The results (\ref{twisteffectA}) and (\ref{twisteffectB})
strongly suggest that sectors paired by the action
of $\Psi_K$ contain states of opposite spacetime statistics,
with the sectors $\tilde A_K$ and $\tilde B_K^{(s)}$
containing states of {\it both}\/ spacetime statistics.
What all of these statististics actually are, though, is largely
irrelevant from the point of view of spacetime physics, for most of
these extra sectors are of course ultimately GSO-projected out
of the fractional-superstring spacetime spectrum in a manner consistent
with modular invariance.
Indeed, only $\tilde A_K$ and $\tilde B_K^{(a)}$ actually
contribute particles to the fractional-superstring spectrum,
and these are presumably bosonic and fermionic.
We shall seek to understand the spacetime statistics of
these states in more detail in Sect.~6.3.

\subsection{The Twist Current as a Compactified-Boson Primary Field}

In Sect.~6.1 the twist current was constructed and analyzed
in terms of the underlying parafermionic CFT's which were the original basis
of the fractional-superstring worldsheet theory.
We have just demonstrated, however, that the action of this current
is a symmetry of the entire {\it post}\/-projection CFT, suggesting
that this twist current is feature of the original parafermionic CFT
which survives the internal projections.
What will interest us here, therefore, is the construction of an operator
directly in the isomorphic post-projection compactified-boson CFT
which has the same effect.  We will find that this
is indeed possible, and that the twist current is in fact isomorphic
to a certain primary field in the compactified-boson CFT.  This
will thereby enable us to interpret the effects of the twist current
on the various $A$- and $B$-sectors in terms of the
properties [such as the fusion rules and $U(1)$ charge
conservation] of the isomorphic compactified-boson theory.
This is of paramount importance if we are to understand the
fractional-superstring post-projection CFT (and
the associated twist current) without reference to its original
parafermionic formulation.

Our line of attack will be first to determine the effect of the
current $\Psi_K$ on the {\it characters} of the compactified-boson theory, and
only subsequently to express this operator directly in terms of the
primary fields of the compactified-boson theory.
Let us first concentrate on the $K=8$ case, which turns out
to be somewhat simpler.
We have already seen in (\ref{UAVeight}) how the three characters
$\tilde A_8$, $U_8$, and $V_8$  of the $A$-sector theory
are related to those of the $N=4$ compactified boson theory,
and in terms of these latter characters,
the action (\ref{twisteffectA}) of the twist current immediately implies:
\beq   \currenteight:~~~~~ \chi_{4,1} ~\leftrightarrow~ -\chi_{4,1},~~~~
                    \chi_{4,0} ~\leftrightarrow~ -\chi_{4,2}~.
\eeq
It is thus evident in this bosonized language
that the action of this current is simply:
\beq  \currenteight:~~~~~  \chi_{N,k} ~\longrightarrow~ -\chi_{N,N/2+k}~
\label{tobegeneralized}
\eeq
where we recall that $\chi_{N,\pm k}=\chi_{N,N+k}$.
However, this form is not quite general enough for our purposes.
For example, recall from (\ref{intrescaling}) that these three
$N=4$ characters could also generally be written as linear combinations
of the $N=4 n^2$ characters, where $n\in \IZ$.
If we had we chosen to write our fractional-superstring characters
$\tilde A_8$, $U_8$, and $V_8$
in terms of the characters of any of these $N=4n^2$ systems for $n>1$,
the above action of $\currenteight$ would not have the desired effect.
We therefore require a general action invariant under any integral
radius rescaling.
The arguments just above (\ref{intrescaling}), however, indicate
that if $R$ is rescaled by a factor of $n$, then $N$ is rescaled
by a factor of $n^2$ but $k$ is rescaled only by a factor of $n$;
indeed, (\ref{intrescaling}) is the statement that for $R\in \IZ$,
the quantities
\beq
      \sum_{p=0}^{\sqrt{N}-1} \,\chi_{N,k+p\sqrt{N}}~,~~~~ 0\leq k<N
\eeq
are invariant under $N\to n^2 N$, $k\to nk$.
Thus, in order for our twist-current action to be invariant under
these radius rescalings, we must modify (\ref{tobegeneralized}):
\beq  \currenteight:~~~~~  \chi_{N,k} ~\longrightarrow~ -\chi_{N,\sqrt{N}+k}~.
\label{modified}
\eeq
This result is consistent with the requirement $k\in\IZ$,
for $\sqrt{N}$ is always an integer for $N=4n^2$.

The above action (\ref{modified}) thus describes the
parafermionic twist current for the $A$-sectors, whose
characters, we recall, can be formulated in terms of $\vartheta$-functions
with unscaled arguments ({\it i.e.}, with scaling parameter $\lambda=1$).
Let us now proceed to the $B$-sector, which for $K=8$ has scaling parameter
$\lambda=5$.  Here our mapping to compactified-boson characters
in (\ref{eightidentities}) proves especially fruitful,
for in terms of these characters
we see that the twist current $\currenteight$ again has the simple action
$\chi_{N,k} \rightarrow -\chi_{N,N/2+k}$.
Rewriting this result so that it is invariant
under integer radius-rescalings,
we therefore have our result for the $B$-sector:
\beq  \currenteight:~~~~~  \chi_{N,k} ~\longrightarrow~ -\chi_{N,\sqrt{5N}+k}~.
\label{modifiedtwo}
\eeq
Thus, it is clear that (\ref{modified}) and (\ref{modifiedtwo})
for both the $A$- and $B$-sectors can
be generally written together in the $K=8$ case as
\beq  \currenteight:~~~~~
       \chi_{N,k} ~\longrightarrow~ -\chi_{N,\sqrt{\lambda N}+k}~.
\label{twisteffecteight}
\eeq
This then reproduces the action of $\currenteight$
at the level of the partition-function characters
in both the $A$- and $B$-sectors, with the minus sign representing
the projection/non-projection sector interchange.

We now turn to the $K=4$ case, which is slightly more difficult
because we are now dealing with {\it bilinears} of
compactified-boson characters.
In the $A$-sector, we know from the identities (\ref{UAVfour})
that our current must transform
\beqn
    \currentfour:&&~~~~~~~ (\chi_{4,1})^2 ~\longleftrightarrow~
             -(\chi_{4,1})^2
       \nonumber\\
   &&~~~~~~~ (\chi_{4,0})^2 +(\chi_{4,2})^2~\longleftrightarrow~
             -2\,\chi_{4,0}\,\chi_{4,2} ~.
\eeqn
Because these expressions are now bilinears of compactified-boson
characters, however, there are two possible means of effecting
these transformations.  The first is via the rule
\beq  \currentfour:~~~~~~
      \chi_{N,k} ~\longrightarrow~
      {1\over \sqrt{2}}\,
    \left( e^{i\pi/4} \,\chi_{N,k} - e^{-i\pi/4}\, \chi_{N,\sqrt{N}-k}\right)~,
\eeq
which is again invariant under integral radius rescalings.
This rule, however, has the undesirable property that phases
are introduced as the coefficients of {\it characters}, and
although this causes no problem for the particular combinations of
character {\it bilinears} that we face, this rule
is clearly not conducive to a general interpretation.
Thus, we instead adopt a second possibility:
\beq  \currentfour:~~~~~
       \chi_{N,k_1}\,\chi_{N,k_2}
      ~\longrightarrow~ -\half\,\left(
     \chi_{N,k_1}\,\chi_{N,\sqrt{N}+k_2} ~+~
       \chi_{N,\sqrt{N}+k_1}\,\chi_{N,k_2}\right)~,
\eeq
which has the advantage that all phases are manifestly avoided
for {\it all}\/ combinations of character bilinears.
Using (\ref{fouridentities}),
it is easy to check that this rule applies for the $B$-sectors of
the $K=4$ theory as well, with $\sqrt{N}$ replaced by $\sqrt{5N}$.
We thus have the general twist-current rule for the $K=4$ case:
\beq  \currentfour:~~~~~
       \chi_{N,k_1}\,\chi_{N,k_2}
      ~\longrightarrow~ -\half\,\left(
     \chi_{N,k_1}\,\chi_{N,\sqrt{\lambda N}+k_2} ~+~
       \chi_{N,\sqrt{\lambda N}+k_1}\,\chi_{N,k_2}\right)~.
\label{twisteffectfour}
\eeq

Having thus rather simply described the action of the currents
in terms of the relevant characters of the compactified-boson theory,
we now turn to the next task:  we would like to mimic the situation
for the twist current in our original parafermionic CFT,
and express this current directly in terms of the underlying primary fields
of the compactified-boson theory.  The above actions for $\current$ could then
be understood as the by-products of the
compactified-boson fusion rules.
This can be done rather straightforwardly.
For the $K=8$ case, we have seen that the general action of the
twist current is given in (\ref{twisteffecteight}),
where the minus sign reflects the exchange of projection and non-projection
sectors.  In the {\it post-projection}\/ theory, however, we do not have
any internal projections remaining, and thus at the level of this underlying
CFT we see that our current must simply transform the primary-field sectors
contributing to $\chi_{N,k}$ into those that contribute to
$\chi_{N,\sqrt{\lambda N}+k}$.  It is a simple matter to determine
the primary fields which effect this change under fusion.
The original sectors are parametrized by their $U(1)$ charges $\alpha$,
and we see from (\ref{alphaconstraint}) that these charges
all satisfy $\sqrt{N}\alpha =k$ (mod $N$).
The sectors contributing to
$\chi_{N,\sqrt{\lambda N}+k}$, however,
have charges $\alpha'$ satisfying $\sqrt{N}\alpha'
= k+\sqrt{\lambda N}$ (mod $N$), and thus in order
to effect this transformation we must increase the charge
of each sector by an amount $\Delta \alpha\equiv\alpha'-\alpha$
satisfying
\beq    \Delta\alpha ~=~ \sqrt{\lambda} ~~({\rm mod}\,N)~.
\eeq
Since (\ref{bosonfusions}) indicates that this $U(1)$ charge
is a conserved quantum number under fusion,
we see that the effect of the current $\currenteight$
can thus be understood simply as the result of fusion with any of
the primary fields $e^{i(p\sqrt{N} +\sqrt{\lambda})\phi}$, $p\in \IZ$.
Choosing the simplest case $p=0$, then, we have the result:
\beq  \currenteight ~=~ \epsilon^{(\sqrt{\lambda})}~=~
    e^{i\sqrt{\lambda} \phi}~.
\label{currenteightresult}
\eeq
(This result is to be interpreted not as a strict equality, of course,
but rather as an isomorphic relation.)
Similar arguments for the $K=4$ case yield the analogous
result in the two-boson system:
\beq  \currentfour ~=~ \half \left(
   \bone_{(1)} \,\otimes\, e^{i\sqrt{\lambda}\phi_2} ~+~
   e^{i\sqrt{\lambda}\phi_1}\,\otimes\, \bone_{(2)}\right)
\label{currentfourresult}
\eeq
where $\bone_{(n)}$ indicates the identity field in the conformal
field theory of the $n^{\rm th}$ compactified boson $\phi_n$.

Given these results,
it is straightforward to verify that
the action of $(\currenteight)^2$ is in accordance
with (\ref{currentsquaredidentity}), for
$(\currenteight)^2$ indeed transforms the set of sectors with
charges satisfying $\sqrt{N}\alpha =k$ (mod $N$) into itself for
each value of $k$.  This follows trivially as a consequence of the
fact that $2\sqrt{\lambda N}=0$ (mod $N$), so that the effective value
of $k$ is increased under $(\currenteight)^2$
by $\Delta k=N$.  According to (\ref{Chidef}),
this increase in $k$ is tantamount to a shift in the one-dimensional
boson momentum lattice
by one full lattice spacing, and thus we see that $\currenteight$ itself
shifts this momentum lattice by
one {\it half}\/-lattice spacing.
This in turn implies that bosonic and fermionic states are
shifted by half-lattice spacings relative to each other, an observation
we shall discuss below.
Note, however, that since (\ref{currenteightresult}) effects a simple lattice
translation $\Delta \alpha$, it squares
to one only by acting on precisely those lattices with
lattice spacing $2\Delta \alpha$.

The $K=4$ case is analogous but more complicated.
The current (\ref{currentfourresult}) also does not manifestly
square to {\bf 1} when acting on arbitrary two-dimensional lattices;
rather, we find
\beq
   (\currentfour)^2~=~
  \quart\left( {\bone}_{(1)}\otimes e^{i2\sqrt{\lambda}\phi_2} +
  e^{i2\sqrt{\lambda}\phi_1}\otimes \bone_{(2)}\right) ~+~
  \half\, e^{i\sqrt{\lambda}\phi_1}\otimes e^{i\sqrt{\lambda}\phi_2}~,
\label{currentfoursquared}
\eeq
and while the first two terms are again simple lattice-preserving
translations, the third term represents a diagonal
lattice translation which in general does not preserve the lattice.
The sectors (\ref{fouridentities})
of our $B$-sector CFT's for $K=4$
consist of only those special two-dimensional lattices for which this
third term {\it is}\/ lattice-preserving, however, and thus we find
$(\currentfour)^2=\bone$ when acting on the entire lattice.

\subsection{Lattices and Spin-Statistics}

We close this section by discussing the relationship between the
momentum lattices appearing in these $K>2$ fractional superstring theories,
and the spacetime statistics of the corresponding states.
We will find a close similarity with the analogous situation
for ordinary $K=2$ superstrings, as well as some crucial differences.
Indeed, these differences will illustrate quite dramatically
that our post-projection theories are ultimately {\it not}\/
equivalent to compactified-boson theories, and that despite the
isomorphism between the two which has been discussed in previous
sections, an alternative non-bosonic representation for our post-projection
CFT's remains to be found.

In the $K=8$ case, we have seen that the twist current $\currenteight$
amounts to a translation of the one-dimensional lattice by a half-lattice
spacing, so that bosonic and fermionic states are shifted by half-lattice
spacings relative to each other.  This observation provides a natural
explanation for the fact, discussed in Sect.~3.3, that the
$\tilde A_8$ and $\tilde B_8\equiv\tilde B_8^{(a)}$
sectors contain states of both bosonic {\it and}\/ fermionic
spacetime statistics.  Indeed,
(\ref{UAVeight}) and (\ref{eightidentities}) tell us that
\beqn
     \tilde A_8 ~&\equiv &~ \half\left( A_8^b +A_8^f\right)~=~
      \eta^{-2}\left( \chi_{4,1} + \chi_{4,-1} \right)\nonumber\\
     \tilde B_8 ~&\equiv &~ \half\left( B_8^b +B_8^f\right)~=~
      \eta^{-2}\left( \chi_{20,5} + \chi_{20,-5} \right)~,
\eeqn
yet we see that in general
the lattices corresponding to $\chi_{N,\sqrt{\lambda N}/2}$ and
$\chi_{N,-\sqrt{\lambda N}/2}$ are shifted by exactly a
half-lattice spacing relative to each other.
Thus, we can separately identify the bosonic and fermionic
sectors with the lattice sites of $U(1)$ charges $\alpha^{b,f}$ satisfying
\beqn
    K=8:~~~~~~~ \alpha^{b}~&=&~ + \half\,\sqrt{\lambda}~~~~~
    ({\rm mod} ~2\sqrt{\lambda})\nonumber\\
                \alpha^{f}~&=&~ - \half\,\sqrt{\lambda}~~~~~
    ({\rm mod} ~2\sqrt{\lambda})
\label{alphabfeight}
\eeqn
while maintaining consistency with the twist current interpretation.
Thus we see that $\tilde A_8$ and $\tilde B_8$ need not each correspond to
a single vacuum state at all:  rather, they each correspond to
an infinite number of vacuum ground states which can naturally be divided
into two classes according to (\ref{alphabfeight}).

A similar situation exists for the $K=4$ case.
Here our identities (\ref{UAVfour}) and (\ref{fouridentities})
tell us that
\beqn
    \tilde A_4 ~\equiv~\half(A_4^b+A_4^f)~&=&~ \eta^{-4}\, \left\lbrack
 \chi_{4,1}^{(1)} \chi_{4,1}^{(2) }+
 \chi_{4,-1}^{(1)} \chi_{4,-1}^{(2) }+
 \chi_{4,1}^{(1)} \chi_{4,-1}^{(2) }+
 \chi_{4,-1}^{(1)} \chi_{4,1}^{(2)  }\right\rbrack~\nonumber\\
    \tilde B_4 ~\equiv~\half(B_4^b+B_4^f)~&=&~ \eta^{-4}\, \left\lbrack
 \chi_{12,3}^{(1)} \chi_{12,3}^{(2) }+
 \chi_{12,-3}^{(1)} \chi_{12,-3}^{(2) }+
 \chi_{12,3}^{(1)} \chi_{12,-3}^{(2) }+
 \chi_{12,-3}^{(1)} \chi_{12,3}^{(2) }\right\rbrack\nonumber\\
\label{stepone}
\eeqn
where the superscripts explicitly indicate the
two dimensions of the momentum lattice.
Under the action of the $K=4$ twist current
(\ref{currentfourresult}), however, we see
that the first two terms and the last two terms in each line of (\ref{stepone})
are transformed into each other.  There thus again exists a natural
boson/fermion identification consistent with the twist current:
\beqn
  K=4:~~~~~~~~~
     \vec \alpha^{b}~&=&~  (\pm \half\sqrt{\lambda},\pm \half\sqrt{\lambda})
         ~~~~~~({\rm mod}~ 2\sqrt{\lambda})\nonumber\\
     \vec \alpha^{f}~&=&~  (\pm \half\sqrt{\lambda},\mp \half\sqrt{\lambda})
         ~~~~~~({\rm mod}~ 2\sqrt{\lambda})~.
\label{alphabffour}
\eeqn

These identifications of bosonic and fermionic states in fact
satisfy a number of other properties reminiscent of so-called ``lattice
strings'' (among which is the ordinary $K=2$ superstring).
As is well-known, lattice strings have states whose two
dimensional (worldsheet) left-moving and right-moving momenta together
form a $[\half(D-2),\half(D-2)]$-dimensional ``shifted Lorentzian lattice''
$\Lambda$:
this means that any state $\vec\alpha=(\vec\alpha^{\rm \,left}|
\vec\alpha^{\rm \,right})$ can be written as $\vec\alpha=\vec L+\vec S$
where $\vec S$ is a constant shift-vector known as a ``statistics vector'',
and where the set of vectors $\vec L$ forms a true lattice $\Lambda_s$ with
inner product $\vec L_1 \cdot \vec L_2\equiv \vec L_1^{\rm \,left}\cdot
\vec L_2^{\rm \,left} - \vec L_1^{\rm \,right}\cdot \vec L_2^{\rm \,right}$.
(The shifted lattice $\Lambda$ is not a true lattice because $\vec \alpha_1+
\vec\alpha_2 \not\in \Lambda$.)
The spacetime spin-statistics of a given state $\vec \alpha \in \Lambda$
are then generally discernible by computing
$(\vec \alpha-\vec S)\cdot \vec S= \vec L\cdot \vec S$:
\beq
        \vec L \cdot \vec S ~\in~ \cases{
           \IZ &  bosonic \cr
           \IZ + 1/2 &  fermionic~. \cr}
\label{LdotS}
\eeq
In this formalism, it is also clear that the general three-point
vertex operator in the light-cone picture must be associated with
the worldsheet momentum $\vec S$, so that in the interaction $1+2\to 3$
the momentum of a final state $\vec \alpha_3$ is given not by the sum
$\vec\alpha_1+\vec\alpha_2 \not\in \Lambda$,
but rather by $\vec\alpha_1+\vec\alpha_2+\vec S\in \Lambda$.

For example, for the ordinary ten-dimensional (Type IIA) superstring,
the $(4+4)$-dimensional lattice formed by the GSO-surviving Ramond
and NS states is
\beq
     \Lambda_{K=2} ~=~\Lambda_L~\otimes \Lambda_R
\label{superstringlattice}
\eeq
where the left- and right-moving lattices are given by
\beq
      \Lambda_L~=~ \Lambda_{R}~=~
    \biggl\lbrace n_1,n_2,n_3,n_4 \biggr\rbrace \,\oplus\,
    \biggl\lbrace n_1-\half,n_2-\half,n_3-\half,n_4-\half \biggr\rbrace
\label{latticeleft}
\eeq
with $n_i\in\IZ$ and $\sum n_i ={\rm odd}$.
The shift vector can then be taken to be
\beq
       \vec S_{K=2} ~=~  (1,0,0,0 ~|~ 1,0,0,0)~,
\label{Mandelstamfactor}
\eeq
whereupon the resulting lattice $\Lambda_2-\vec S$ is a true lattice, with
the spacetime bosonic and fermionic states $\vec\alpha$
respectively distinguished according to
(\ref{LdotS}).  From this we see that the chiral twist current $\Psi_2$
of the ordinary superstring takes the simple form
\beq
       \Psi_2 ~=~
        e^{i\sqrt{\lambda}\phi^+_1/2} \otimes
        e^{i\sqrt{\lambda}\phi^+_2/2} \otimes
        e^{i\sqrt{\lambda}\phi^+_3/2} \otimes
        e^{i\sqrt{\lambda}\phi^+_4/2}
\label{Psitwo}
\eeq
where we are now explicitly indicating the chirality of the
bosons $\phi_i$ (and
where we of course have only an ``$A$-sector'' with $\lambda=1$);
this result is to be compared with (\ref{currenteightresult}) and
(\ref{currentfourresult}), where a similar chirality is understood.
Indeed, the partition function (\ref{superpartf})
of the ordinary superstring can now be recast in the usual fashion
as a sum over lattice sites:
\beq
   {\cal Z}_2 ~=~  {\tau_2}^{-4} \, |\eta|^{-24}\,
     \sum_{\vec \alpha \in \Lambda_2}
    q^{(\vec \alpha^{\rm \,left})^2/2}
     \,\overline q^{(\vec \alpha^{\rm \,right})^2/2} \,
     \exp\left\lbrack  2\pi i \,(\vec\alpha -\vec S)\cdot \vec S\right\rbrack~,
\label{Zrecast}
\eeq
and each three-point vertex is associated with
an operator contributing the worldsheet momentum (\ref{Mandelstamfactor}).
The necessity for such a momentum
insertion for the superstring in light-cone gauge
was first perceived by Mandelstam in the early 1970's \cite{Mandelstam},
and indeed this operator contributes the conformal dimension $h=1/2$
(that of the worldsheet supercharge) needed
to produce a dimension-one vertex operator for consistent string emission.

For the fractional superstrings, the statistics assignments
(\ref{alphabfeight})
and (\ref{alphabffour}) permit a similar lattice interpretation.
In particular, for the $A$-sectors
({\it i.e.}, for $\lambda=1$),
these states together fill out the $(8/K+ 8/K)$-dimensional shifted lattice
\beqn
        K>2:  ~~~~~\,\Lambda_K~\equiv~
  \left\lbrace n_1\pm \half,...,n_{8/K}\pm\half \right\rbrace
      \,\otimes\,
  \left\lbrace n_1\pm \half,...,n_{8/K}\pm\half \right\rbrace~,
\eeqn
where $n_i\in \IZ$ and where each sign is chosen independently.
We can then simply take the shift vectors
\beqn
      \vec S_{K=4} ~&=&~ (\half,\half~|~\half,\half)\nonumber\\
      \vec S_{K=8} ~&=&~ (\half~|~\half)~,
\label{shiftsfoureight}
\eeqn
whereupon the resulting set
of states $\Lambda_s\equiv \lbrace \vec\alpha -\vec S\rbrace$
forms a true Lorentzian lattice, and
the assignments (\ref{alphabfeight}) and (\ref{alphabffour}) now follow
directly as a consequence of (\ref{LdotS}).
The $B$-sectors of these strings also have similar lattice
descriptions, of course, with the relevant lattices
scaled by a factor of $\sqrt{\lambda}$.
Thus we can write the fractional-superstring partition functions
for each value of $K$ in a manner analogous to (\ref{Zrecast}):
\beqn
   |A_K|^2 ~&=&~  4\,|\eta|^{-48/K}\,
     \sum_{\vec \alpha \in \Lambda_K}
    q^{(\vec \alpha^{\rm \,left})^2/2}
     \,\overline q^{(\vec \alpha^{\rm \,right})^2/2} \,
     \exp\left\lbrack  2\pi i \,(\vec\alpha -\vec S)\cdot \vec S\right\rbrack~
    \nonumber\\
   |B_K|^2 ~&=&~  4\,|\eta|^{-48/K}\,
     \sum_{\vec \alpha \in \sqrt{\lambda}\Lambda_K}
    q^{(\vec \alpha^{\rm \,left})^2/2}
     \,\overline q^{(\vec \alpha^{\rm \,right})^2/2} \,
     \exp\left\lbrack  2\pi i \,(\vec\alpha/\sqrt{\lambda} -\vec S)\cdot
     \vec S\right\rbrack~.\nonumber\\~
\label{latticerewrite}
\eeqn
This result thus naturally incorporates spacetime supersymmetry,
the internal projections, the spin-statistics assignments, and
the actions of the twist currents in a natural manner.
Indeed, these results suggest that
the analogous fractional-superstring three-point vertex operator
in light-cone gauge should correspond to a worldsheet
momentum insertion $\vec S_K$ with conformal dimension $h=1/K$.
Note that with such an insertion, the spacetime
statistics factor $(-1)^F$ is indeed a conserved quantum
number in all three-point interactions, further suggesting that
our assignments (\ref{alphabfeight}) and (\ref{alphabffour})
are correct.

This analogy between the superstring lattice and these
fractional-superstring lattices must not be pushed too far, however;
fractional superstrings are {\it not}\/ equivalent to
lattice strings, and cannot be described in terms of free worldsheet bosons
via a simple lattice-type construction.
The simplest illustration of this is the fact that the heuristic
fractional-superstring lattices described above are
not {\it self-dual}\/;  this failure follows
as a trivial consequence of the rescaling of the $B$-sector lattices.
Although the self-duality of an underlying lattice implies
invariance of the partition function under the $S$ modular transformation,
it is certainly not the case
that any $S$-invariant partition function corresponds to an underlying
self-dual lattice:  non-trivial counter-examples include orbifold-compactified
string theories, which have no underlying lattice formulations.
The lattice description presented here is thus meant only
to demonstrate the consistency of our twist-current
assignments (\ref{alphabfeight}) and (\ref{alphabffour}), and
their similarity to those of the ordinary superstring.

Indeed, this lattice description also illustrates quite dramatically why the
fractional-superstring post-projection CFT's cannot ultimately
be represented in terms of
free bosons.  This isomorphism between our post-projection CFT's and those of
free bosons has thus far lead us to a lattice description in which each
surviving fractional-superstring state
of highest weight $h$ is associated with a lattice site $\vec\alpha$ satisfying
$h=\vec\alpha\cdot\vec\alpha/2$
in such a way that the fusion rules of the post-projection
CFT are equivalent to vector
addition for $\vec\alpha$.  Our spacetime statistics assignments
(\ref{alphabfeight}) and (\ref{alphabffour}) are then the only ones consistent
with the twist current, and manifestly imply spacetime supersymmetry.
However, for $K>2$ we cannot take the next step, and actually associate each
lattice site $\vec\alpha$ with the worldsheet boson field
$\exp(i\vec\alpha\cdot \vec\phi)$.  The problem may be seen as follows.
For the ordinary superstring, (chiral) lattice sites
$\vec\alpha\in\Lambda_{L,R}$ generally
correspond to spacetime bosons or fermions depending on whether their lattice
coordinates $\alpha_i$ are integer or half-integer.
This is consistent with their interpretation as arising from the fields
$\exp(i\vec\alpha\cdot \vec\phi)$ in the free-boson theory, for each of the
primary fields $\exp(\pm i\phi_i/2)$ is equivalent to a tensor product of two
Ising-model spin fields $\sigma$, and these spin fields create the necessary
worldsheet cuts to alter the boundary conditions of worldsheet fermions and
produce fermionic spacetime statistics.  For $K>2$, however, the statistics
assignments in (\ref{alphabfeight}) and (\ref{alphabffour}) clearly preclude
any such free-boson representation;  indeed, only the fermionic states in the
$A$-sectors appear representable in this manner.
Therefore an alternative representation
for our light-cone worldsheet theory is needed, one which is consistent not
only with these spacetime statistics assignments, but more generally with
transverse $(D_c-2)$-dimensional spacetime Lorentz invariance.\footnote{
  Note that there are in fact two states at each lattice site,
  so that the massless states ({\it i.e.}, those with $\vec\alpha^2=2/K$)
  indeed have the correct multiplicities to fill out
  $SO(D_c-2)$ vector and spinor representations.}
Presumably the new massive
$B$-sector states (with their additional rescaled lattice
sites) will be important in this regard.

\section{Summary and Concluding Remarks}
\setcounter{footnote}{0}

In this paper we have taken the first steps towards a construction
of the post-projection worldsheet conformal field theory of the fractional
superstring for both the $A$-sectors and the $B$-sectors.  By explicitly
demonstrating how the internal projections rearrange the original
parafermionic Fock space of the fractional superstring at the level of the
corresponding characters, we have determined the central charges, the
complete spectrum of highest weights, and the corresponding fusion rules of
these post-projection conformal field theories:  for the $A$-sectors these
results are summarized in (\ref{Aresultstwo}) with the characters of the
$A$-subsectors given in (\ref{Abf}), (\ref{Avacids}), and (\ref{moreids});
and for the $B$-sectors the characters and highest weights are given in
(\ref{Bsectorsfour}) and (\ref{Bsectorseight}), with the fusion rules given
in (\ref{Bfusionsfour}) and (\ref{Bfusionseight}).   We then demonstrated
that all of these results have a natural reformulation in terms of worldsheet
compactified bosons [Eqs.~(\ref{UAVeight})--(\ref{fouridentities})],
and this reformulation ultimately enabled us not only to
demonstrate that the twist current
is a symmetry of the entire post-projection CFT for both the $A$- and
$B$-sectors [Eqs.~(\ref{twisteffectA}) and (\ref{twisteffectB})],
but also to construct this current directly in terms of the primary fields
of the post-projection CFT's [Eqs.~(\ref{currenteightresult}) and
(\ref{currentfourresult})].  This in turn permitted us to
find a consistent separation of spacetime states into those that
are spacetime bosonic or fermionic for both the $A$- and $B$-sectors
[Eqs.~(\ref{alphabfeight}) and (\ref{alphabffour})], and
to subsequently express our fractional-superstring
partition functions in lattice language [Eq.~(\ref{latticerewrite})]
in a manner consistent with spacetime supersymmetry, the internal projections,
and the twist-current spacetime statistics interpretation.
Our results also exposed some of the outstanding problems that confront a
``bottom-up'' construction of suitable representations for
these post-projection CFT's, and taken together
they can therefore be viewed as providing a set of
minimal constraints that any future construction must satisfy.

As indicated at the end of Sect.~5.3, however, one of the results of our
analysis is that the $A$- and $B$-sectors after the internal projections
appear to be described as fully independent CFT's with completely uncoupled
fusion rules.  A natural question then arises as to whether the
$A$- and $B$-sector post-projection CFT's can be described together
as different subsectors of a single, larger post-projection CFT with central
charge $c=24/K$;  indeed, we generally expect there to be interactions
or couplings between $A$- and $B$-sector states, and these interactions
should emerge naturally in a properly formulated larger CFT.
We shall therefore conclude by briefly describing several
different ideas with might ultimately prove useful
in determining this single, larger light-cone CFT.  We note, however, that it
is of course conceivable that no such single worldsheet CFT exists for
the light-cone version of the fractional superstring, and that the passage
from a covariant formulation to light-cone gauge contains new
features rendering the $A$- and $B$-sector subtheories apparently separate.

The first idea involves the compactification of bosons on orbifolds.
We have seen in (\ref{radii}) that the $A$-sector post-projection CFT's
can be formulated as tensor products of bosons compactified on circles
of radius $R_{A}=1$, while the $B$-sectors correspond to products
of bosons compactified on circles of radius $R_{B}=\sqrt{\lambda} =
\sqrt{\half(K+2)}$.  Thus, any single CFT which is to contain these theories
as separate subtheories must simultaneously have characters corresponding
to bosons of radius $R_{A}$ as well as characters corresponding
to a different radius $R_B$.  One group of theories which has this property is
that of a boson compactified on an orbifold.  For example, the partition
function of a single boson compactified on a $\IZ_2$ orbifold of general
radius $R$ is given by
\beq
   Z_{\rm orb}(R)~=~ \half \,\left[
    Z_{\rm circ}(R) ~+~ 2\,Z_{\rm circ}(\sqrt{2}) ~-~ Z_{\rm circ}(1/\sqrt{2})
    \right]~
\label{Zorb}
\eeq
where $Z_{\rm circ}(R)$ is the partition function of the circle-compactified
boson given in (\ref{ZofR}) and (\ref{ZofRtwo}).  In (\ref{Zorb}), the first
term is the contribution from the {\it untwisted}\/ sector, while
\beq
   Z_{\rm twist} ~\equiv~
    Z_{\rm circ}(\sqrt{2}) - \half\,Z_{\rm circ}(1/\sqrt{2}) ~=~
  \half\,|\chi_{8,0}-\chi_{8,4}|^2 + 2 \,|\chi_{8,1}|^2 + 2 \,|\chi_{8,3}|^2
\label{Ztwist}
\eeq
is the contribution from the $\IZ_2$-twisted sector.  Other more complicated
types of orbifold theories have partition functions similar to
these \cite{Ginsparg}.   Thus, we see that
the use of orbifolds allows us to effectively combine circle-compactified
theories of different radii, at least at the character (or partition-function)
level.  Furthermore, note that the untwisted sector has a
radius which equals the radius of the orbifold, while the effective
radii of the twisted sectors are constants which turn out to depend only on
the type of orbifold being considered.  This suggests that the
post-projection CFT of the $K$-fractional superstring might
consist of an $8/K$-fold tensor product of $c=1$ orbifold theories at radius
$R_{\rm orb}=\sqrt{\half(K+2)}$
(along with the $D_c-2=16/K$ original coordinate bosons):
the $B$-sectors would then correspond to the {\it untwisted}\/
orbifold sectors, and the $A$-sectors would correspond to the {\it twisted}\/
orbifold sectors.  This is certainly consistent with the observation
that the $A$-sector is always a tensor product
of the {\it same} $R=1$ theory for each value of $K$,
whereas the radius of the $B$-sector theory varies with $K$.
Indeed, for the case of the $\IZ_2$ orbifold, we find
that the contribution $Z_{\rm twist}$
in (\ref{Ztwist}) can be rewritten in terms of {\it unscaled}\/
Jacobi $\vartheta$-functions regardless of the value of $R_{\rm orb}$:
\beq
     Z_{\rm twist} ~=~  \half\, |\eta|^{-2}\,\biggl(
     |\vartheta_2\vartheta_3|+
     |\vartheta_2\vartheta_4|+
     |\vartheta_3\vartheta_4| \biggr)~.
\label{Ztwisttheta}
\eeq

The problem with this $\IZ_2$ orbifold approach, however, is readily
apparent from (\ref{Ztwisttheta}):  the $\IZ_2$ orbifold theory
does not contain the proper ``diagonal'' characters  $\chi_{h}^{(c=8/K)}$
which (as we have found) comprise the $A$-sector CFT.
Similar difficulties arise for the other types of $c=1$ orbifold theories
as well.  Therefore, since these circle- and orbifold-compactified theories
have been proven to completely span the space of unitary $c=1$
CFT's \cite{Ginsparg,Kiritsis}, these observations indicate that this
$c=8/K$ component of the desired unified CFT will not
be a simple $8/K$-fold tensor product of $c=1$ theories.

Another character-based approach\footnote{
          This approach was developed in collaboration with P.C.~Argyres.}
towards determining the desired larger CFT can be formulated by
avoiding the assumption that these $c=8/K$ component theories can be
represented
as tensor-products of $c=1$ theories, and by instead working directly with the
$A$- and $B$-sector characters determined in Sects.~3 and 4.
Recall that we found a total of
twelve such characters for each value of $K$:  these were the
$A$-sector characters $\lbrace U_K, \tilde A_K, V_K\rbrace$, as well as the
$B$-sector characters $\lbrace X_K^{(s)}, \tilde B_K^{(s)}, Y_K^{(s)}\rbrace$
for $s=a,b,c$.  While each of these sets of characters is separately closed
under modular transformations, our goal is of course to construct a larger
set of characters which not only contains these two smaller sets but which
also introduces a non-trivial mixing between them.
(In the language of Sect.~5.3, this is tantamount to constructing a
new set of larger representation matrices $S_{ij}$ and $T_{ij}$ so that
the $A$- and $B$-sectors are {\it both}\/ necessary for a consistent
partition function.)   We already have one
indication of how these sectors should mix, however.  Recall that the original
linear combinations $A_K \equiv A_K^b-A_K^f$ and $B_K \equiv B_K^b-B_K^f$
[which are ``orthogonal'' to $\tilde A_K\equiv \half(A_K^b+A_K^f)$ and
$\tilde B_K\equiv \half(B_K^b+B_K^f)$]
close into each other under modular transformations ---
indeed, it is due to such mixings that the $B$-sectors originally appeared
in the modular-invariant partition functions (\ref{partfuncts}).
This suggests that our twelve-member character set should be
enlarged through the addition of the two extra linear combinations $A_K$ and
$B_K$, or equivalently by treating the four characters $A_K^{b,f}$ and
$B_K^{b,f}$ as completely independent.  This is certainly consistent with our
spacetime statistics assignments, according to which these four characters
correspond to separate and distinguishable Fock spaces.  Moreover,
this method of introducing a non-trivial coupling between the $A$- and
$B$-sectors does not require the introduction of further additional sectors,
since the resulting set of fourteen characters is again closed
under the modular group.  However, the fundamental problem one encounters in
attempting to interpret these fourteen characters as those of a single CFT
is the presence of {\it two}\/ vacuum characters, $U_K$ and
$X_K^{(a)}$, both with $h=0$.  Presumably some linear combination of these
characters corresponds to the true vacuum state of the unified single CFT
(assuming that such a CFT exists), but a quick check demonstrates that there
do not appear to be any such combinations which simultaneously satisfy a
number of other physical criteria.

A third approach to this problem might be to avoid the characters
altogether, and to start with the bosonic lattice formulation presented in
Sect.~6.3:  we would then attempt to non-trivially combine the lattice of
$A$-sector states with the rescaled lattice of $B$-sector states.  However,
since the $B$-sector lattice scaling factor $\sqrt{\lambda}$ is irrational,
the $A$- and $B$-sector lattices are incommensurate, and their direct sum
does not form a lattice (with or without a shift vector).  Indeed, filling
out the remaining lattice sites necessary to form a true lattice yields a
set of lattice points which is dense in each lattice direction, and the
physics of such a ``lattice'' is unclear.  Alternatively, one might
assume a coupling between the $A$-sector and $B$-sector theories which
takes place only through
the states corresponding to the origin $\vec\alpha=0$, for this is the one
lattice point which these theories have in common.  In this scenario,
then, $A$-sector states and $B$-sector states would be coupled
only if the sums of the internal momenta of the $A$- and $B$-sector states
each separately vanish.  Unfortunately, this type of restricted coupling
is highly unusual, and certainly has no analogue in the ordinary superstring.
Whether a consistent theory can be formulated with such a restricted coupling
in light-cone gauge remains to be seen.

Finally, regardless of whether there exists a single unified light-cone CFT
for the fractional superstring, there of course remains the issue we have
faced from the beginning:  that of properly constructing our worldsheet CFT(s)
in terms of worldsheet fields in a manner consistent with spacetime
$(D_c-2)$-dimensional Lorentz invariance and proper spacetime statistics.
While our results concerning the twist current provide valuable clues as to
what the statistics of these CFT sectors are expected to be, such results
should emerge naturally in a proper formulation of the worldsheet CFT.
In particular, this entails finding a representation of our light-cone
CFT's so that the states in our various sectors have well-defined
transformation properties under the transverse Lorentz group $SO(D_c-2)$.

Thus, our results concerning the post-projection worldsheet conformal field
theories of the fractional superstring constitute only the first steps in their
eventual construction, and many issues remain to be resolved before the
consistency of the fractional superstring --- both on the worldsheet and in
spacetime --- is demonstrated.  Work in all of these areas is continuing.

\bigskip
\medskip
\leftline{\large\bf Acknowledgments}
\medskip

We are pleased to thank Philip Argyres, C.S.~Lam, and Henry Tye
for many useful discussions, and we acknowledge the hospitality of
the Aspen Center for Physics where portions of this work were initiated.
This work was supported in part by the Natural Sciences and Engineering
Research Council of Canada and by les fonds FCAR du Qu\'ebec.

\bigskip

\bibliographystyle{unsrt}

\end{document}